\documentclass[lettersize,journal]{IEEEtran}

\usepackage{amsmath,amsfonts}
\usepackage{algorithmic}
\usepackage{algorithm}
\usepackage{array}
\usepackage[caption=false,font=normalsize,labelfont=sf,textfont=sf]{subfig}
\usepackage{textcomp}
\usepackage{stfloats}
\usepackage{url}
\usepackage{verbatim}
\usepackage{graphicx}
\usepackage{cite}
\usepackage{xcolor}
\usepackage{xr}
\usepackage{multirow}
\usepackage{cleveref}
\usepackage{xcolor}
\usepackage{soul}
\usepackage{color, colortbl}
\usepackage{geometry}
\usepackage{amsmath}
\usepackage{mathtools}
\usepackage{booktabs}

\begin{document}

\title{
Can a Machine Feel Vibrations?: A Framework for Vibrotactile Sensation and Emotion Prediction via a Neural Network
}


\author{Chungman Lim, Gyeongdeok Kim, Su-Yeon Kang, Hasti Seifi, and Gunhyuk Park

\thanks{Manuscript received December 13, 2024; revised XX XX, 2025.}}

\markboth{IEEE TRANSACTIONS ON HAPTICS,~Vol.~XX, No.~X, August~2025}%
{Lim \MakeLowercase{\textit{et al.}}: A Sample Article Using IEEEtran.cls for IEEE Journals}


\maketitle

\begin{abstract}
Vibrotactile signals offer new possibilities for conveying sensations and emotions in various applications. 
Yet, designing vibrotactile tactile icons (i.e., Tactons) to evoke specific feelings often requires a trial-and-error process and user studies. 
To support haptic design, we propose a framework for predicting sensory and emotional ratings from vibration signals.
We created 154 Tactons and conducted a study to collect acceleration data from smartphones and roughness, valence, and arousal user ratings (n=36).
We converted the Tacton signals into two-channel spectrograms reflecting the spectral sensitivities of mechanoreceptors, then input them into VibNet, our dual-stream neural network.
The first stream captures sequential features using recurrent networks, while the second captures temporal-spectral features using 2D convolutional networks.
VibNet outperformed baseline models, with 82\% of its predictions falling within the standard deviations of ground truth user ratings for two new Tacton sets.
We discuss the efficacy of our mechanoreceptive processing and dual-stream neural network and present future research directions.
\end{abstract}


\section{Introduction}
Tactile icons (i.e., Tactons) play a crucial role in a variety of haptic devices, such as wearable devices~\cite{friesen2023perceived, schirmer2015shoe}, VR/AR controllers~\cite{lee2017exploring, seo2018substituting, pezent2019tasbi}, and smartphones~\cite{seo2013perceptual}.
They serve a wide range of functions, from alerting users to events or system states~\cite{gaffary2018use, ryu2010vibrotactile} to mediating sensations and emotions~\cite{brown2007tactons, turchet2012haptic}, thus creating immersive interactions and enhancing user experiences.
Moreover, Tactons support diverse user scenarios, such as communicating emotions with others~\cite{ju2021haptic}, sending emotional messages to children with autism~\cite{changeon2012tactile}, and expressing feelings between long-distance partners~\cite{park2013roles}.

Various design approaches have sought to create effective Tactons for conveying and communicating sensations and emotions. 
Common approaches include parameter-based design~\cite{yoo2015emotional, seifi2013first}, which systematically varies the design parameters of a vibrotactile signal, and metaphor-based design, which creates Tactons that evoke specific metaphors in users~\cite{seifi2015vibviz, seifi2017exploiting}.
For both design approaches, designers need to run multiple experiments with many users to assess subjective sensations and emotions elicited by Tactons.
Through these user studies, designers identify effective design parameters for delivering targeted sensations and emotions, using the findings to create Tactons suitable for specific applications.

Despite the usefulness of these approaches, they come with high costs and effort to conduct user studies, which limit designers in testing new Tactons. 
For instance, if a designer wants to create a new Tacton with a complex rhythmic structure that has not been previously investigated, additional user studies are required to evaluate the sensations and emotions elicited by the new Tacton.
While some guidelines exist on the effects of design parameters on sensations or emotions (e.g., the effect of frequency on roughness~\cite{tan1999information}), designers resort to repetitive trial-and-errors and extensive user testing until they find a Tacton that conveys the desired sensations and emotions to users.

To address these challenges, we ask:
Can we predict the sensations and emotions users feel from Tactons using vibration accelerations produced by a vibrotactile actuator in haptic devices to accelerate the progress of Tacton design?
In this paper, we introduce a framework for predicting the sensory and emotional ratings of Tactons with three components: (1) haptic data augmentation, (2) mechanoreceptive processing, and (3) a neural network designed to mimic human tactile perception and cognition mechanisms.
To construct a haptic dataset of accelerations paired with corresponding ratings (roughness, valence, and arousal; on a scale of 0 to 100) of Tactons, we first designed 154 Tactons that cover a wide range of design parameters, including carrier frequency, envelope frequency, duration, amplitude, rhythmic structure, and complex waveforms.
We then conducted a user study with 36 participants using three different iPhone models with different masses and sizes to reflect the use of commercial haptic devices.
Next, we augmented the dataset by devising a set of haptic data augmentation methods, considering factors related to human vibrotactile perception, such as the Just Noticeable Difference (JND) for vibrations.
We then processed the augmented data into two-channel spectrograms using biomimetic mechanoreceptive channel filters based on human vibrotactile receptors (Meissner and Pacinian Corpuscles) and Short-Time Fourier Transform (STFT).

We trained a neural network, VibNet, using the accelerations and two-channel spectrograms, structured as two parallel input streams.
The first stream feeds 1D acceleration data into Gated Recurrent Units (GRUs)~\cite{cho2014learning}, a type of recurrent neural networks, to capture sequential features of Tactons.
This approach is grounded in human vibrotactile perception of temporal and rhythmic patterns~\cite{lim2023can, park2011perceptual, ternes2008designing, seifi2013first, yoo2015emotional}.
The second stream feeds the two-channel 2D spectrograms into Convolutional Neural Networks (CNNs), specifically ResNet~\cite{he2016deep}, to capture temporal-spectral features perceived by vibration-relevant mechanoreceptive channels in the skin~\cite{choi2012vibrotactile, lederman2009haptic}.
Through these two streams, VibNet predicts roughness, valence, and arousal ratings, which are the primary sensory and emotional dimensions for Tactons.

We evaluated the trained model using two new Tacton sets based on~\cite{lim4785071emotional}, each containing 24 Tactons designed by varying signal parameters and 24 complex Tactons.
Our model demonstrated state-of-the-art performance compared to baseline machine learning models, such as Long Short-Term Memory (LSTM) and Transformer, achieving lower root mean square errors (RMSE) between the predictions and the ground truth ratings.
On average, 82\% of the predictions generated by our model fell within the standard deviation of the ground truth ratings across the three dimensions of sensation and emotion for all 48 Tactons in the two test sets.
Based on these results, we present the implications of our framework and discuss directions for future research on predicting vibrotactile sensations and emotions.
Our contributions include:
\begin{itemize}
\item Sensory and emotional ratings for 154 Tactons varying in six design parameters, and corresponding 1D accelerations measured from three commercial devices.
\item A biomimetic framework including haptic data augmentation, mechanoreceptive processing, and a neural network (VibNet), with an end-to-end implementation.
\item A demonstration of the efficacy of our method in predicting the sensation and emotion evoked by new unseen Tactons.
\end{itemize}


\section{Related Work}
We review prior research on design parameters for Tactons in haptics, studies on vibrotactile perception, sensation, and emotion, and computational models for haptic stimuli.

\subsection{Design Parameters and Libraries for Tactons}
Over the decades, extensive research have explored myriad design parameters for Tactons to effectively convey and communicate information and emotions in various haptic user interfaces~\cite{maclean2008foundations}.
Prior studies have investigated low-level parameters of Tactons, such as carrier frequency~\cite{hwang2010perceptual, israr2006frequency}, envelope frequency~\cite{park2011perceptual, lim2023can}, duration~\cite{kwon2023can, yoo2015emotional}, and amplitude~\cite{hwang2010perceptual, israr2006frequency}, as well as superposition of multiple sinusoids~\cite{yoo2022perceived, hwang2017perceptual} and combinations of multiple parameters~\cite{yoo2015emotional, lim2023can}.
Other studies have proposed high-level parameters of Tactons, such as rhythmic structure~\cite{ternes2008designing, brown2006multidimensional, abou2022vibrotactile}, which include the evenness of pulses and note length, interval between vibrations~\cite{tan2019user}, and sound waveform or timbre~\cite{brown2007tactons, brown2005first}.
Designers can create Tactons by systematically varying these parameters (i.e., parameter-based design approach).
For constructing a haptic dataset, we select four common low-level Tacton parameters -- carrier frequency, envelope frequency, duration, and amplitude -- and one high-level parameter, rhythmic structure.

In addition to creating Tactons from scratch, haptic designers can create new Tacotns by transforming libraries from other modalities into vibration libraries or by modifying template Tactons from existing vibration libraries~\cite{schneider2016studying}.
Past studies have proposed various vibration libraries together with their associated user feelings or metaphors (i.e., metaphor-based design approach).
For example, van Erp and M.A. Spap\'e created 59 Tactons by transforming auditory melodies into vibrations and examined their perceptual impacts~\cite{van2003distilling}.
Disney Research introduced FeelEffects, a library of over 40 Tactons, and investigated the semantic and parametric spaces of these Tactons~\cite{israr2014feel}.
Seifi et al. proposed VibViz, a library of 120 Tactons with subjective ratings and descriptive tags on their physical, sensory, emotional, usage, and metaphoric attributes~\cite{seifi2015vibviz}.
These libraries consist of Tactons with complex waveforms, such as intricate rhythmic structures in the time domain, varying frequency spectra over time, and various durations.
We include 40 Tactons by modifying existing Tactons from the open-source vibration library VibViz for rendering on iPhones, to enhance the diversity of our haptic dataset.

\subsection{Studies on Vibrotactile Perception, Sensation, and Emotion}

Investigating how humans perceive vibrations is a fundamental aspect of the haptics field.
Prior research has studied the neurophysiological processes underlying tactile perception, identifying four types of mechanoreceptors in human skin that contribute to the perception of touch~\cite{goldstein1989sensation}.
These four mechanoreceptive tactile channels have different characteristics, such as perception properties, spectral sensitivities, and sensory adaptation rates~\cite{kandel2000principles, choi2012vibrotactile}.
Among these four mechanoreceptive channels, Meissner Corpuscle (RA1) and Pacinian Corpuscle (RA2) are primarily activated by vibrotactile stimuli compared to Merkel Disk (SA1) and Ruffini Ending (SA2).
Drawing on the properties of the two mechanoreceptive channels (RA1 and RA2) most related to vibrations and their spectral sensitivities, we propose a mechanoreceptive processing approach that converts vibration signals into two-channel spectrograms.

Previous studies have also explored the perceptibility and discriminability of vibrations.
Psychophysics researchers have examined the detection thresholds, or Absolute Limens (AL), for vibrations and uncovered a U-shaped threshold curve across frequencies, with the highest sensitivity occurring around 200\,Hz (between 150\,Hz and 300\,Hz)~\cite{gescheider2013psychophysics, ryu2010psychophysical}.
Other studies have investigated the discrimination thresholds, or Just Noticeable Differences (JND), between vibrations~\cite{israr2006frequency, franzen1975vibrotactile, goble1994vibrotactile, goff1967differential}.
JNDs for vibration intensity generally range from 10\% to 30\%, while JNDs for vibration frequency typically fall between 15\% and 30\%~\cite{choi2012vibrotactile}.
Some studies have also explored the relative impact of these parameters; increasing the duration of vibrations has been shown to decrease the JNDs for vibration intensity~\cite{gescheider1996effects, jones2006human}.
We use these established findings on detection and discrimination thresholds for vibrations to augment our vibration signals while ensuring that the augmented vibrations remain perceptually indistinguishable to users.

The sensations and emotions elicited by haptic stimuli are also important research topics in HCI.
Past research has developed sensory and emotional lexicons associated with haptic stimuli~\cite{holliins1993perceptual, hollins2000individual, soufflet2004comparison, tiest2006analysis}.
Researchers have collected user ratings based on Russell's circumplex model of affect, which defines two key emotional dimensions: valence and arousal~\cite{russell1980circumplex}.
Guest et al. identified 26 sensory attributes and 14 emotional attributes for describing touch and demonstrated that valence and arousal are primary factors in haptic emotional experience~\cite{guest2011development}.
Among sensory attributes, researchers identified roughness as the primary dimension for capturing the perceptual properties of real textures or materials, with low variability among individual users.
Studies on Tactons reported analogous findings: participants associated Tactons with roughness effectively but often faced challenges in associating hardness with Tactons, and sensory attributes like wet/dry or hot/cold were not relevant to vibrations~\cite{seifi2013first, seifi2015vibviz}.
Additionally, while temporal attributes such as tempo, energy, and rhythm are relevant for describing certain vibration patterns, they are not considered primary sensory dimensions, as they are context-dependent and less generalizable across different tactile scenarios.
Building on these established frameworks, we select roughness as the primary sensory dimension for Tactons and valence and arousal as the two primary emotional dimensions.

To inform the design of Tactons that convey specific sensations and emotions, previous studies have sought to derive sensory and emotional spaces for Tactons and identify effective design parameters.
After creating one or more sets of Tactons, designers typically recruit participants to gather subjective responses to these Tactons.
The sensory and emotional spaces are then visualized using averaged ratings for each Tacton, or statistical tests are applied to determine which design parameters significantly impact the sensations and emotions.
Seifi and Maclean designed 14 Tactons and demonstrated that rhythmic structure influenced all three dimensions (roughness, valence, and arousal) and that carrier frequency affected arousal~\cite{seifi2013first}.
Yoo et al. explored emotional spaces with three different Tacton sets (25, 36, and 24 patterns) and provided design guidelines for four sinusoidal parameters~\cite{yoo2015emotional}.
Seifi et al. collected roughness, valence, and arousal ratings for 120 Tactons and developed the VibViz visualization to assist haptic designers~\cite{seifi2015vibviz}.
While these studies have examined emotional and sensory spaces through controlled laboratory user studies (i.e., offline studies), recent research has explored the efficacy of crowdsourcing user studies (i.e., online studies) to reduce the effort and time needed to collect sensory and emotional ratings of Tactons~\cite{schneider2016hapturk, lim4785071emotional}. 
However, these studies still rely on the efforts of researchers and designers in evaluating Tactons, which limits their scalability.
To address this limitation, we construct sensory and emotional ratings of 154 Tactons, the largest Tacton set ever studied, and develop a model that predicts roughness, valence, and arousal ratings to enable the rapid prototyping of effective Tacton candidates.

\subsection{Computational Models for Haptic Stimuli}

Computational models that predict subjective evaluations of haptic stimuli or compare different haptic stimuli can help designers and researchers enhance user experiences and accelerate the design process in various applications. 
Previous research has developed models to predict haptic perceptions and sensations for objects or textured surfaces.
These studies used robots or rigid tools to collect haptic data generated from physical interactions, such as accelerations, forces, and speeds.
The resulting models predict sensory attributes elicited from objects~\cite{chu2015robotic, richardson2020learning, gao2016deep}, textured surfaces~\cite{awan2023predicting, ito2021model}, or perceptual similarities between textured surfaces~\cite{richardson2022learning}.
While these models primarily focus on real objects or textured surfaces, we propose a model that predicts the sensations and emotions elicited by Tactons conveyed through a single vibrotactile actuator in haptic interfaces, which generates 1D accelerations.

Past studies have also proposed computational models for predicting subjective evaluations for vibrations rendered on a vibrotactile actuator.
Park and Kuchenbecker introduced algorithms to transform three-axes accelerations collected from human interactions with textured surfaces into one-axis accelerations that can be played on a vibrotactile actuator~\cite{park2019objective}.
They then developed a model to assess the perceptual similarities between the haptic stimuli generated from human interactions with textured surfaces and the converted one-axis accelerations.
Others researchers proposed models for evaluating and comparing perceptual qualities between original and compressed vibration signals~\cite{hassen2019subjective, muschter2021perceptual, noll2022automated}.
Recent work introduced a model to predict perceptual dissimilarities between Tactons by simulating the neural transmission from mechanoreceptors in the skin to the brain~\cite{lim2023can}.
Previous studies primarily focused on developing models to predict perceptual distinguishability between real-world haptic stimuli and vibrations or between different Tactons.
In contrast, we propose a computational model that predicts sensations and emotions elicited by Tactons.
We note that unless perceptual differences between Tactons are extremely subtle (i.e., Tactons are not easily distinguishable), users may perceive their sensory and emotional attributes differently. 
This highlights the need for vibrotactile sensation and emotion prediction models, as perceptual distinguishability models alone may not capture the nuanced mappings between vibrations and user-rated sensation and emotions.
Our approach enables efficient prediction for new and unique Tactons that may not closely resemble existing vibrations in the dataset.

\begin{figure*}[t]
  \centering
    \includegraphics[width=\textwidth]{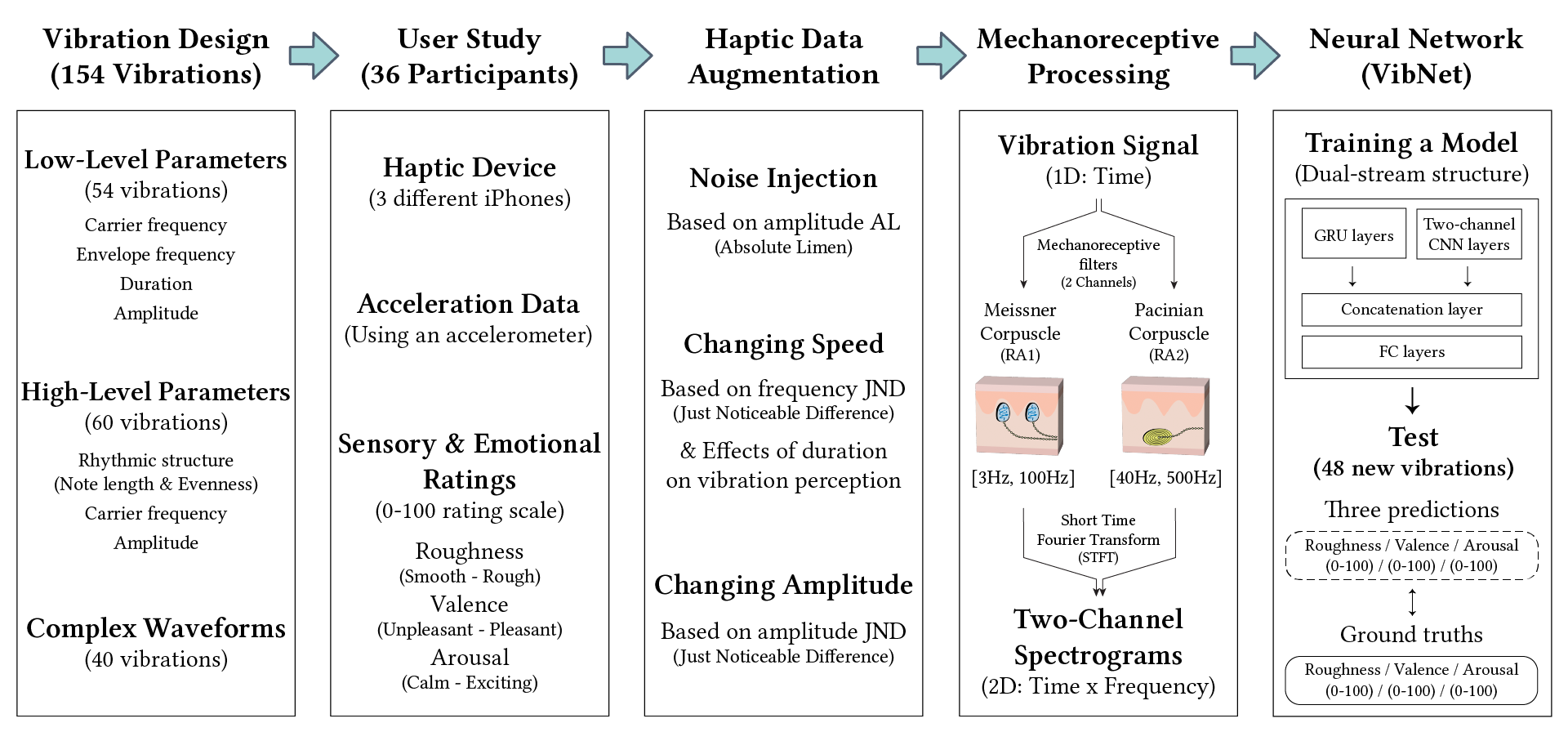}
    \caption{
    An overview diagram illustrating the Tacton design, user study to construct a haptic dataset, and our computational framework.
    }
  \label{fig:overview}
\end{figure*}

\section{Haptic Dataset Construction}
\label{sec:DataConstruction}

We conducted a user study to record acceleration data from 154 Tactons rendered on three commercial devices (iPhones) and to collect sensory and emotional ratings of the Tactons (Figure~\ref{fig:overview}).

\subsection{Tacton Design}

\begin{figure*}[t]
  \centering

    \subfloat[]{
    \centering
    \includegraphics[width=0.31\linewidth]{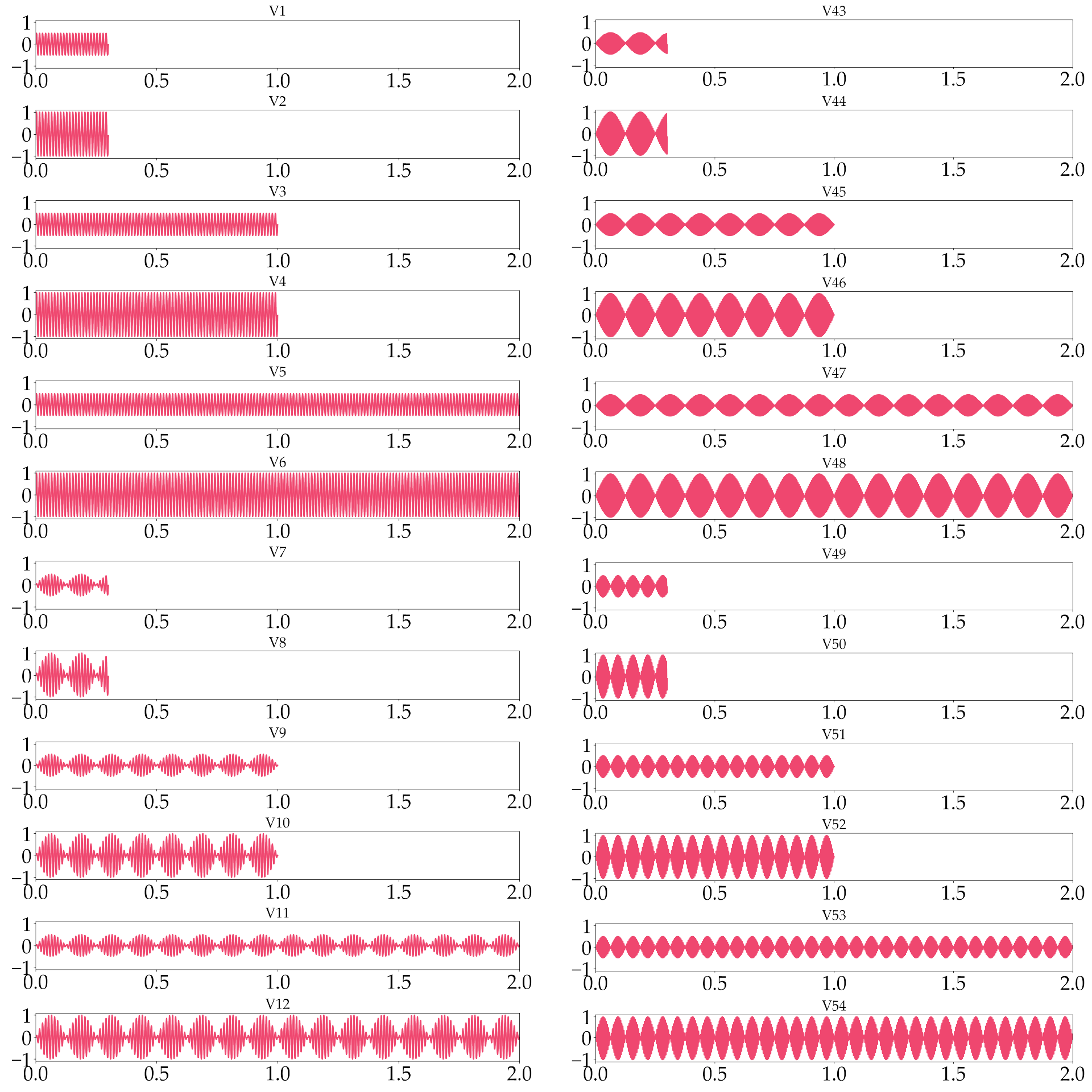}
    \label{fig:v-sinusoid}
    }
    \hfill
    \subfloat[]{
    \centering
    \includegraphics[width=0.31\linewidth]{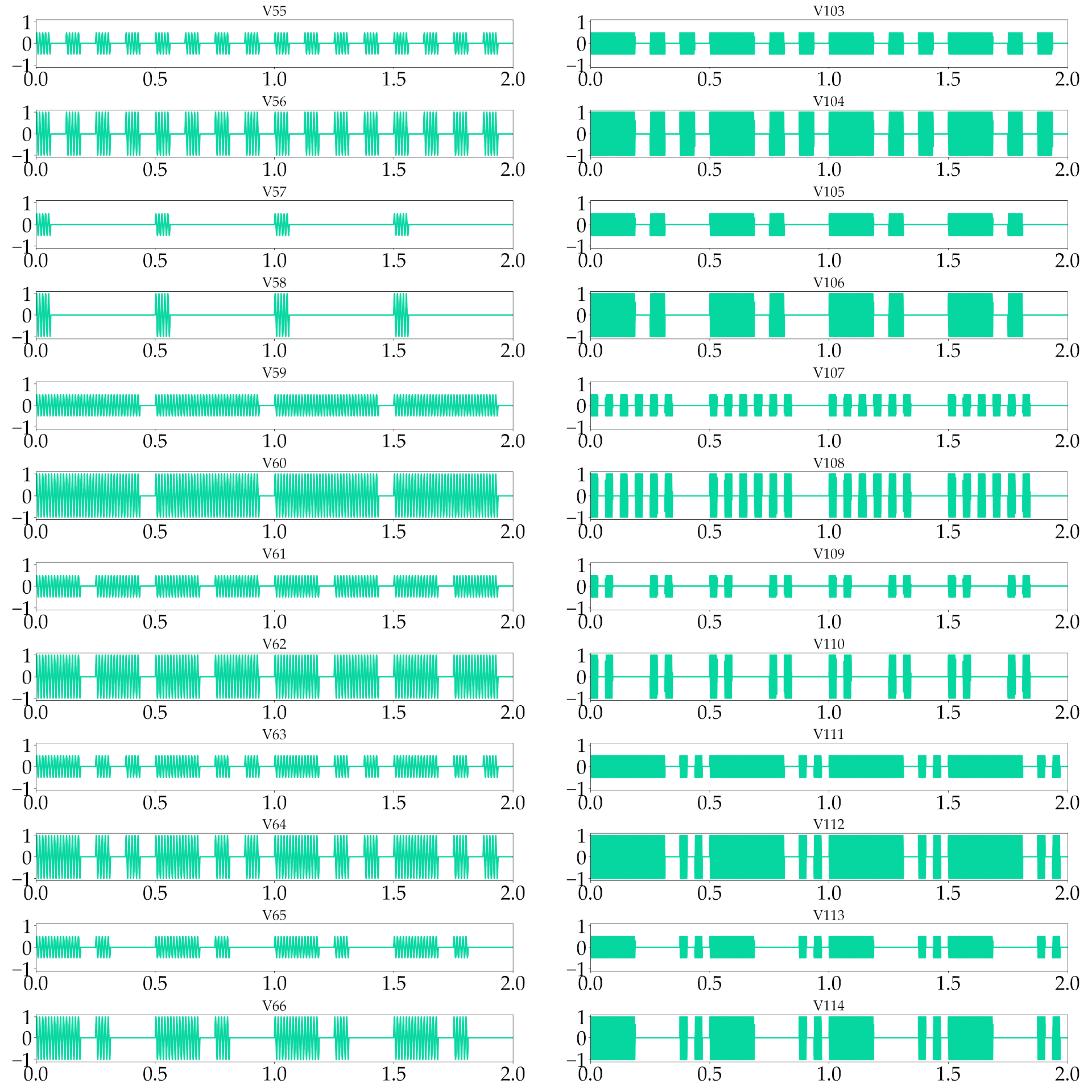}
    \label{fig:v-rhythm}
    }
    \hfill
    \subfloat[]{
    \centering
    \includegraphics[width=0.31\linewidth]{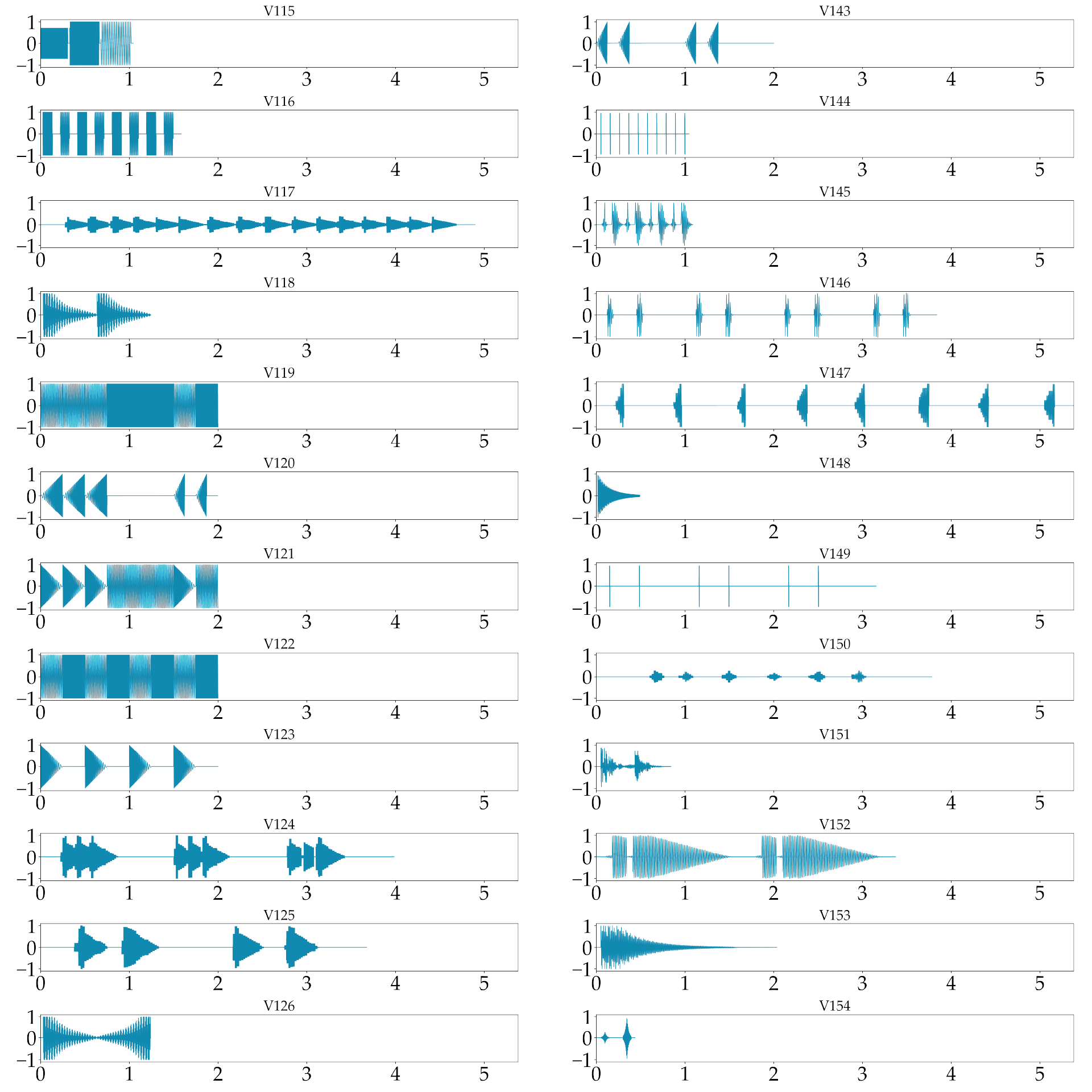}
    \label{fig:v-complex}
    }

    \caption{ 
    Examples of 154 Tactons created using three design approaches:
    (a) 24 examples from 54 Tactons with varying sinusoidal parameters,
    (b) 24 examples from 60 Tactons with varying rhythmic structure, carrier frequency, and amplitude, and
    (c) 24 examples from 40 Tactons created based on VibViz~\cite{seifi2015vibviz}.
    The x-axis shows time (seconds), and the y-axis shows amplitude (-1 to 1) on iPhones. 
    }
  \label{fig:vibrationDesign}
\end{figure*}

We created a set of 154 Tactons to provide haptic dataset to the neural networks (Figure~\ref{fig:vibrationDesign}).
The design of this set was informed by prior literature on Tacton design, utilizing sinusoidal parameters (54 Tactons), rhythmic structures (60 Tactons), and complex Tactons that evoke multiple attributes such as sensations, emotions, and metaphors (40 Tactons).

The 54 Tactons (V1--V54) systematically varied in signal parameters, including three carrier frequencies (80\,Hz, 155\,Hz, and 230\,Hz), three envelope frequencies (0\,Hz, 4\,Hz, and 8\,Hz), three durations (300\,ms, 1000\,ms, and 2000\,ms), and two amplitudes ($half$ and $full$) (Figure~\ref{fig:v-sinusoid}).
We selected these parameters to cover most of the emotional space of sinusoidal vibrations~\cite{yoo2015emotional} while ensuring compatibility with iPhone playback.
We rendered the sinusoidal vibrations using the mathematical formula for temporal envelopes $E(t)$ and temporal frequencies $F(t)$ for generating Tactons on iPhones, as follows:

\begin{equation}
    \begin{dcases}
        E(t) = A \cdot |sin(2 \pi f_{e} t)| \\
        F(t) = f_{c}
    \end{dcases}
    \label{equ:temporalConfiguration_1}
\end{equation}

Here, $A$ represents the amplitude in the Apple Haptic and Audio Pattern (AHAP) format, where $half$ and $full$ correspond to 0.5 and 1, respectively.
$f_{e}$ denotes the envelope frequency, where $f_{e} = 0$\,Hz represents a constant envelope (i.e., $E(t) = A$), and $f_{c}$ denotes the carrier frequency.

The 60 Tactons (V55--V114) varied on 10 rhythmic structures from~\cite{ternes2008designing} and two signal parameters of three carrier frequencies (80\,Hz, 150\,Hz, and 230\,Hz) and two amplitudes ($half$ and $full$) (Figure~\ref{fig:v-rhythm}).
We selected these rhythmic structures to cover most of the perceptual space of rhythmic Tactons, considering primary parameters such as note length and evenness.
All 60 Tactons lasted for 2 seconds.
We rendered the rhythmic Tactons using the following formula for temporal envelopes $E(t)$ and temporal frequencies $F(t)$:

\begin{equation}
    \begin{dcases}
        E(t) = A \cdot R(t) \\
        F(t) = f_{c}
    \end{dcases}
    \label{equ:temporalConfiguration_1}
\end{equation}

Here, $A$ and $f_{c}$ represent the amplitude in the AHAP format and the carrier frequency, respectively.
$R_{t}$ is a list of binary pulses, each with a length of 31.25\,ms, as shown in~\cite{abou2022vibrotactile, lim2024designing}.

Lastly, we designed 40 complex Tactons (V115--V154) using the Vibviz library~\cite{seifi2015vibviz} (Figure~\ref{fig:v-complex}).
We divided the original emotional space of Tactons in the library into nine segments and selected 40 Tactons based on their distribution in the original sensory and emotional spaces.
We rendered the Tactons by transforming original vibration waveforms into temporal envelopes $E(t)$ and temporal frequencies $F(t)$ suitable for iPhones.
These Tactons featured more complex waveforms compared to the 54 sinusoidal and 60 rhythmic Tactons.
In addition, these complex Tactons used variable $F(t)$ over time, whereas the 54 sinusoidal and 60 rhythmic Tactons used constant frequency (i.e., $F(t)=f_{c}$).
The durations of the 40 Tactons ranged from 0.43 to 5.38 seconds.

\subsection{Participants}
We recruited 36 participants (18 women and 18 men; 18–31 years old (mean: 22.8, SD: 3.5)), including three left-handed and 33 right-handed users.
All participants had no sensory impairments in both hands.
The participants took 65 minutes on average to complete the study and received \$30 USD as compensation.

\subsection{Experiment Setup}
\label{sec:setup}

\begin{figure*}[t]
  \centering

    \hfill
    \subfloat[]{
    \centering
    \includegraphics[height=4.5cm]{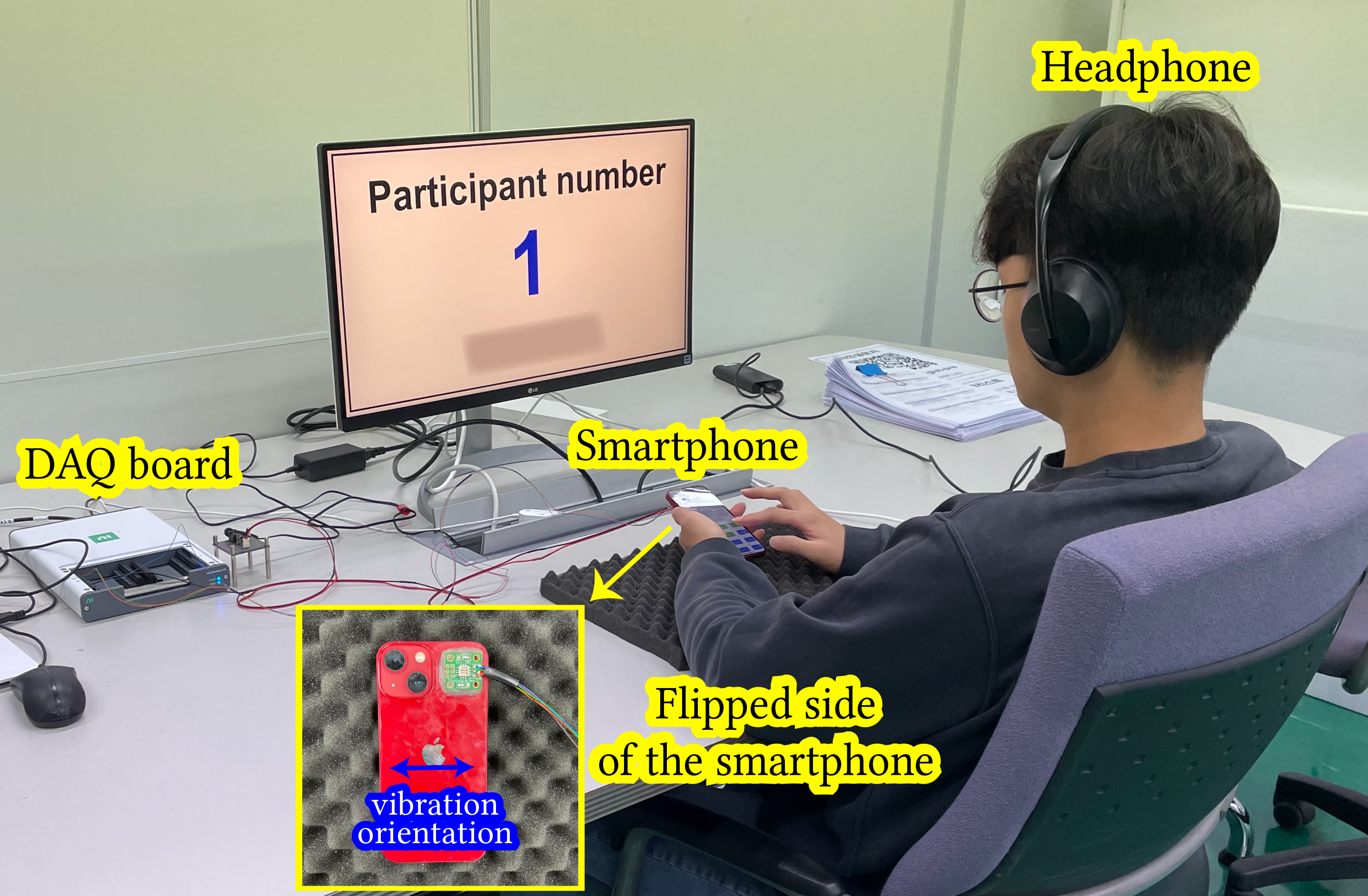}
    \label{fig:setup}
    }
    \hfill
    \hfill
    \subfloat[]{
    \centering
    \includegraphics[height=4.5cm]{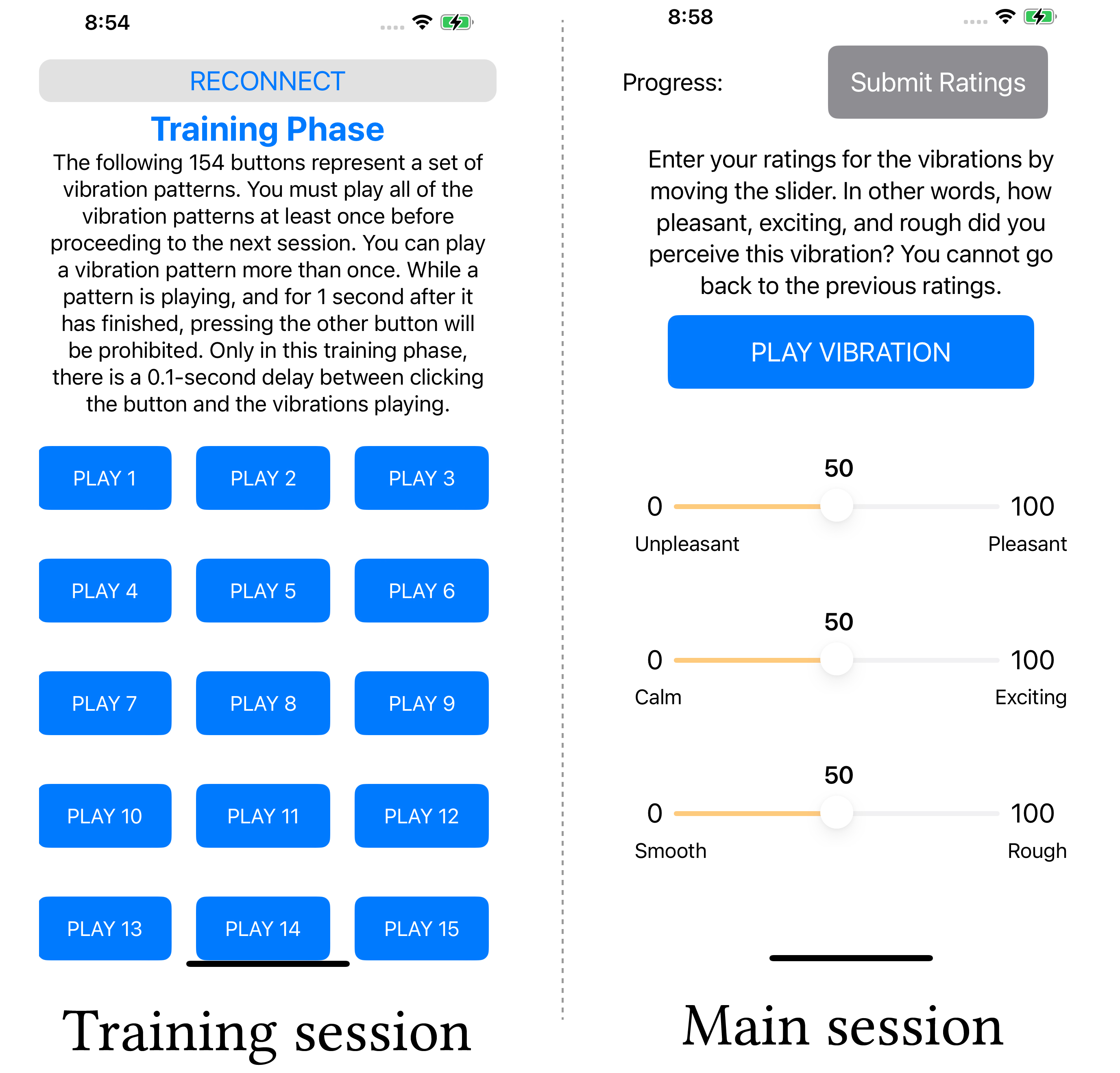}
    \label{fig:screenshot}
    }
    \hfill
    \hfill

  \caption{
  Study setup: (a) An overview of the experimental setup and (b) A screenshot of the GUI application used to collect user responses in the user study (left: training session, right: main session).
  }
  \label{fig:Experiment_Setup}
\end{figure*}

We used three iOS smartphones (iPhone 13 mini, iPhone 14, and iPhone 11 Pro Max) by Apple Inc. to collect acceleration data as well as sensory and emotional ratings of Tactons on various consumer phones.
These phones varied in size and mass: iPhone 13 mini ($64.2 \times 131.5 \times 7.65$,mm, 141,g), iPhone 14 ($71.5 \times 146.7 \times 7.8$,mm, 172,g), and iPhone 11 Pro Max ($77.8 \times 158.0 \times 8.1$,mm, 226,g).
We attached a 3-axis accelerometer (Analog Devices; ADXL354z) on the right to the smartphone camera (Figure~\ref{fig:setup}) and measured the accelerations of the vibrations using a DAQ board (National Instruments; USB-6353) with the sample rate of 10\,kHz. 
We collected three-axis accelerations but used one axis (left-right) aligned with the vibration direction of the Taptic Engines in the iOS smartphones when training a neural network, as the other two axes consisted of noise unrelated to the Tactons.
We collected participants' responses using a graphical user interface (GUI) on the smartphones (Figure~\ref{fig:screenshot}).
Participants placed both hands on a table in front of them and maintained the same holding posture with their left hand and a natural grasp force while interacting with the GUI application using their right hand.
The participants wore noise-canceling headphones with white noise to block any environmental sounds.

\subsection{Experiment Procedure}
\label{sec:ExperiementProcedure}

The study process included a sequence of sessions: introduction, training session, and main session.
Participants first completed the consent form and a demographics pre-questionnaire.
Next, participants began the experiment using the GUI application on one of three smartphones, with participants evenly distributed by biological sex across the devices.
The training session displayed a set of buttons, each randomly corresponding to a Tacton (Figure~\ref{fig:screenshot} left).
Participants experienced all the Tactons before the main session, while we collected acceleration data from their interactions using the accelerometer.
If a participant played a Tacton multiple times, the acceleration data from the last interaction was stored.
In other words, we measured the accelerations from 12 participants (six women, six men) per smartphone, resulting in acceleration data for 154 Tactons from 36 participants across the three smartphones(Figure~\ref{fig:accelerations}).

In the main session, participants rated roughness (smooth/rough), valence (unpleasant/pleasant), and arousal (calm/exciting) of 154 Tactons using sliders ranging from 0 to 100 (Figure~\ref{fig:screenshot} right).
The application presented Tactons in a random order.
Participants could play the Tactons multiple times but were unable to modify ratings for previously rated Tactons.
Throughout the study, participants could take breaks as needed, and we maintained the room temperature between 20--23 degrees Celsius.

\subsection{Results}

We collected 5,544 acceleration data (154 Tactons $\times$ 36 participants) and downsampled all the data from a 10\,kHz to a 1\,kHz, a sample rate that captures the range of human tactile receptors.
This was done before augmenting the vibration signals and inputting them into the neural network to improve the efficiency of model training.

\begin{figure*}[t]
  \centering

    \subfloat[]{
    \centering
    \includegraphics[width=0.98\linewidth]{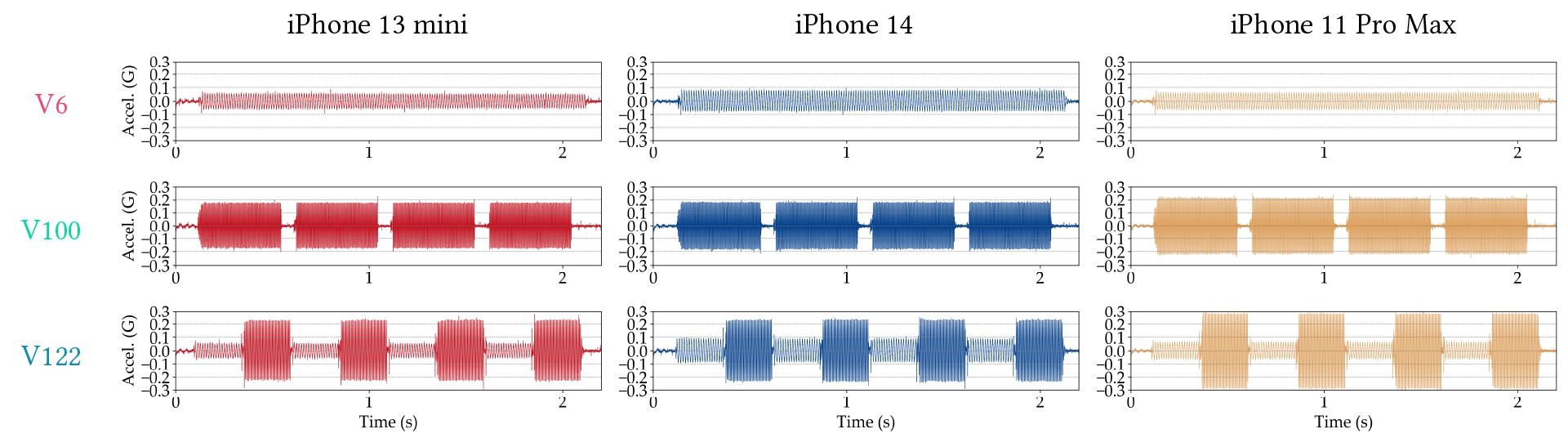}
    \label{fig:accelerations}
    }
    \hfill
    
    \hfill
    \subfloat[]{
    \centering
    \includegraphics[width=0.96\linewidth]{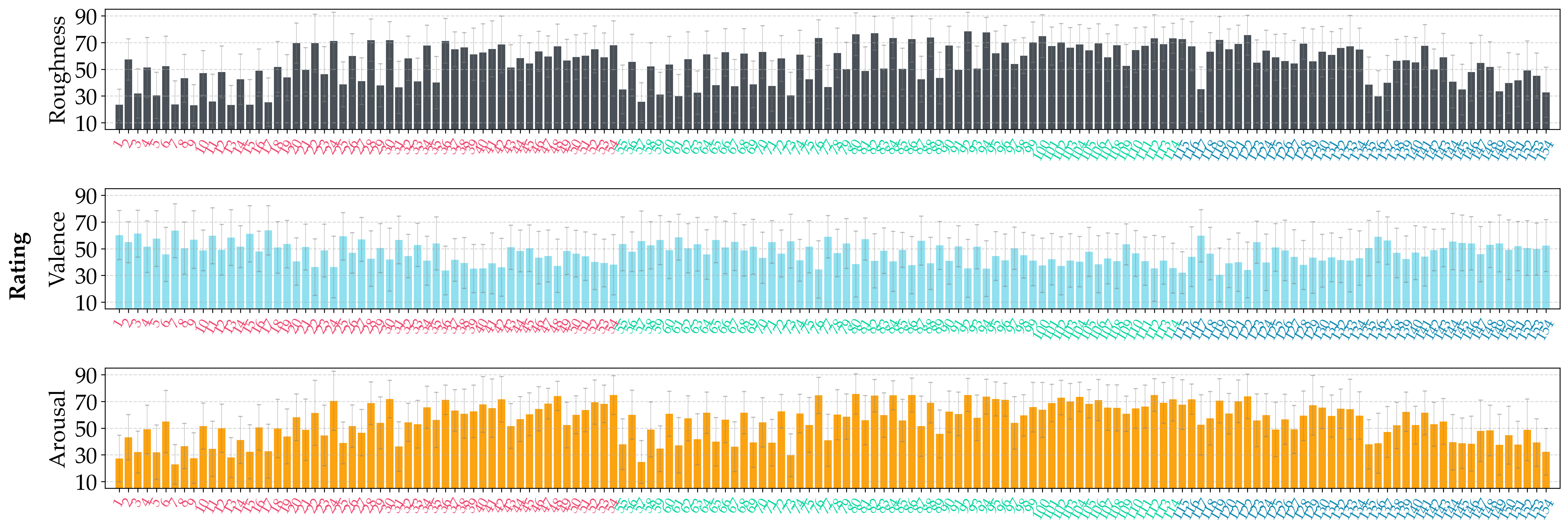}
    \label{fig:resultsEach}
    }

    \caption{
    Accelerations and sensory and emotional ratings of 154 Tactons collected from 36 participants.
    (a) Exemplar accelerations of three Tactons (V6, V100, and V122) from the three devices (iPhone 13 mini, iPhone 14, and iPhone 11 Pro Max).
    (b) The average ratings and standard deviations for roughness, valence, and arousal of 154 Tactons.
    }
  \label{fig:studyResults}
\end{figure*}

In the comparisons in ratings between the three devices (averaged for 12 participants per device), roughness and arousal ratings showed very strong correlations.
For roughness: Pearson's correlation $r$ = 0.91 (iPhone 14 vs. iPhone 13 mini), 0.90 (iPhone 14 vs. iPhone 11 Pro Max), and 0.90 (iPhone 13 mini vs. iPhone 11 Pro Max), all with $p < 0.01$.
For arousal: $r$ = 0.87 (iPhone 14 vs. iPhone 13 mini), 0.88 (iPhone 14 vs. iPhone 11 Pro Max), and 0.88 (iPhone 13 mini vs. iPhone 11 Pro Max), all with $p < 0.01$.
In contrast, valence ratings showed low to moderate correlations between the iPhone 14 and the other two devices ($r$ = 0.38, $p < 0.01$ for iPhone 13 mini and $r$ = 0.47, $p < 0.01$ for iPhone 11 Pro Max), while the correlation in valence ratings between the iPhone 13 mini and iPhone 11 Pro Max was strong ($r$ = 0.72, $p < 0.01$).
Therefore, we used the averaged ratings for each device to train the neural network.
Overall, valence ratings showed a negative relationship with arousal ratings (mean $r = -0.79$ across the three devices), while arousal ratings had a positive relationship with roughness ratings (mean $r = 0.89$), as shown in~\cite{seifi2015vibviz}.
The standard deviations for roughness, valence, and arousal ratings of the 154 Tactons across 36 participants were 17.2, 19.1, and 17.3 (out of 100), respectively (Figure~\ref{fig:resultsEach}).

\section{Haptic Data Augmentation and Mechanoreceptive Processing}
In this section, we augment the 5,544 acceleration data entries, as increasing the scale of data can help improve the performance of neural networks, reduce overfitting, and enhance generalization of the model~\cite{goodfellow2016deep}.
Next, we introduce a mechanoreceptive processing technique for converting 1D accelerations into two-channel 2D spectrograms to feed the processed data into our proposed neural network.

\subsection{Haptic Data Augmentation}
\begin{figure*}[t]
  \centering

    \subfloat[Noise Injection]{
    \centering
    \includegraphics[width=0.3\linewidth]{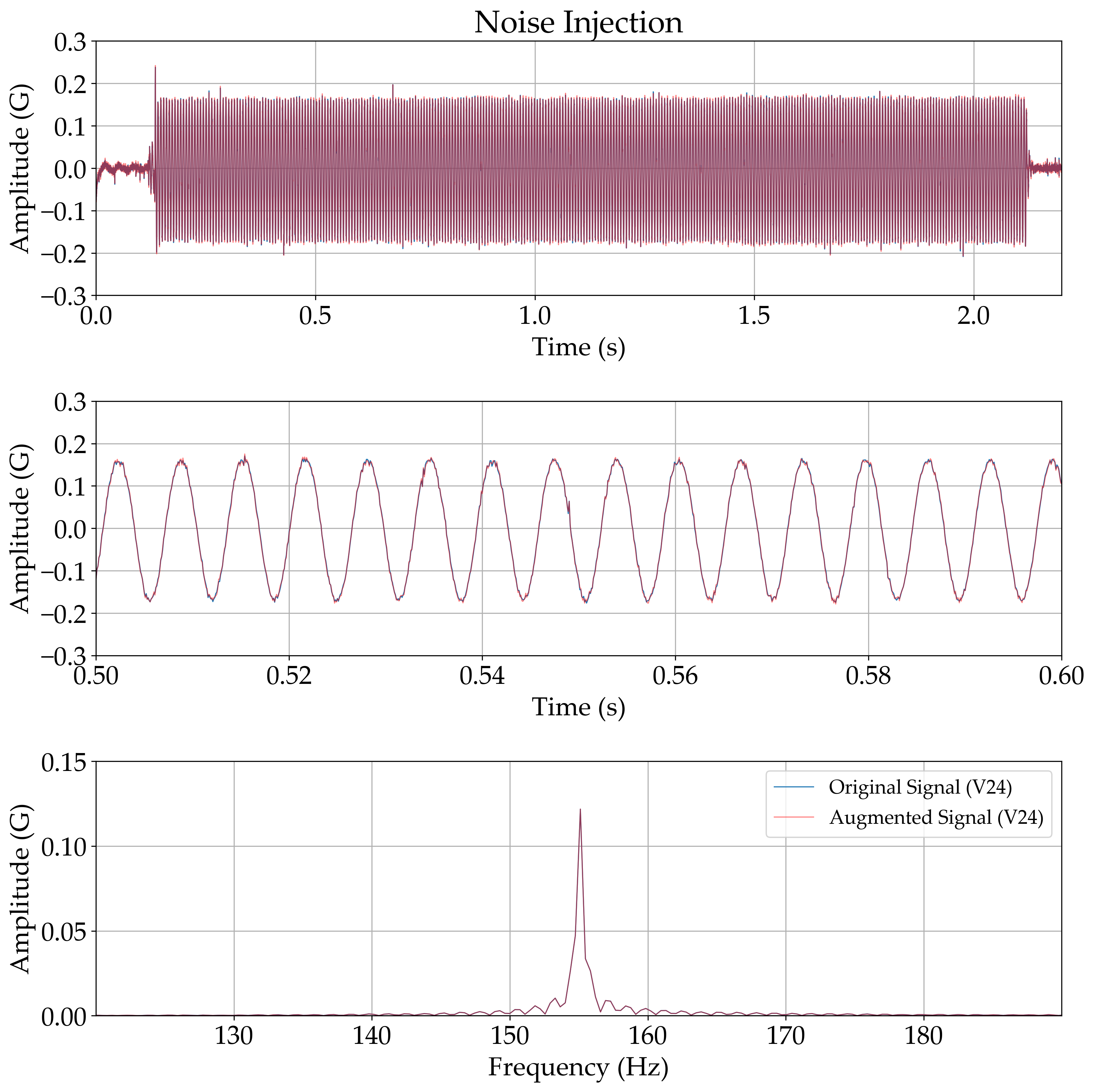}
    \label{fig:aug-noise}
    }
    \hfill
    \subfloat[Changing Speed]{
    \centering
    \includegraphics[width=0.3\linewidth]{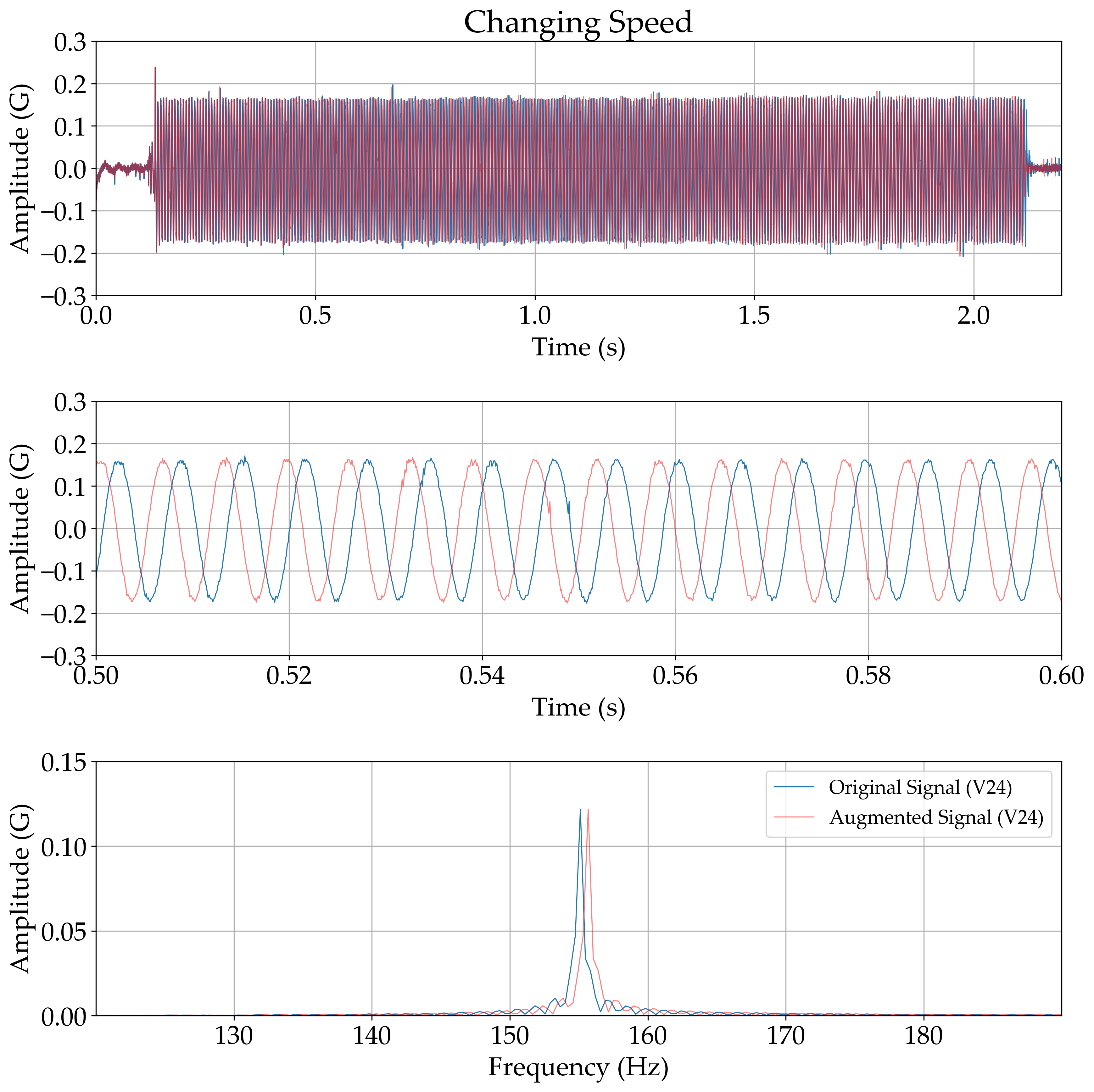}
    \label{fig:aug-frequency}
    }
    \hfill
    \subfloat[Changing Amplitude]{
    \centering
    \includegraphics[width=0.3\linewidth]{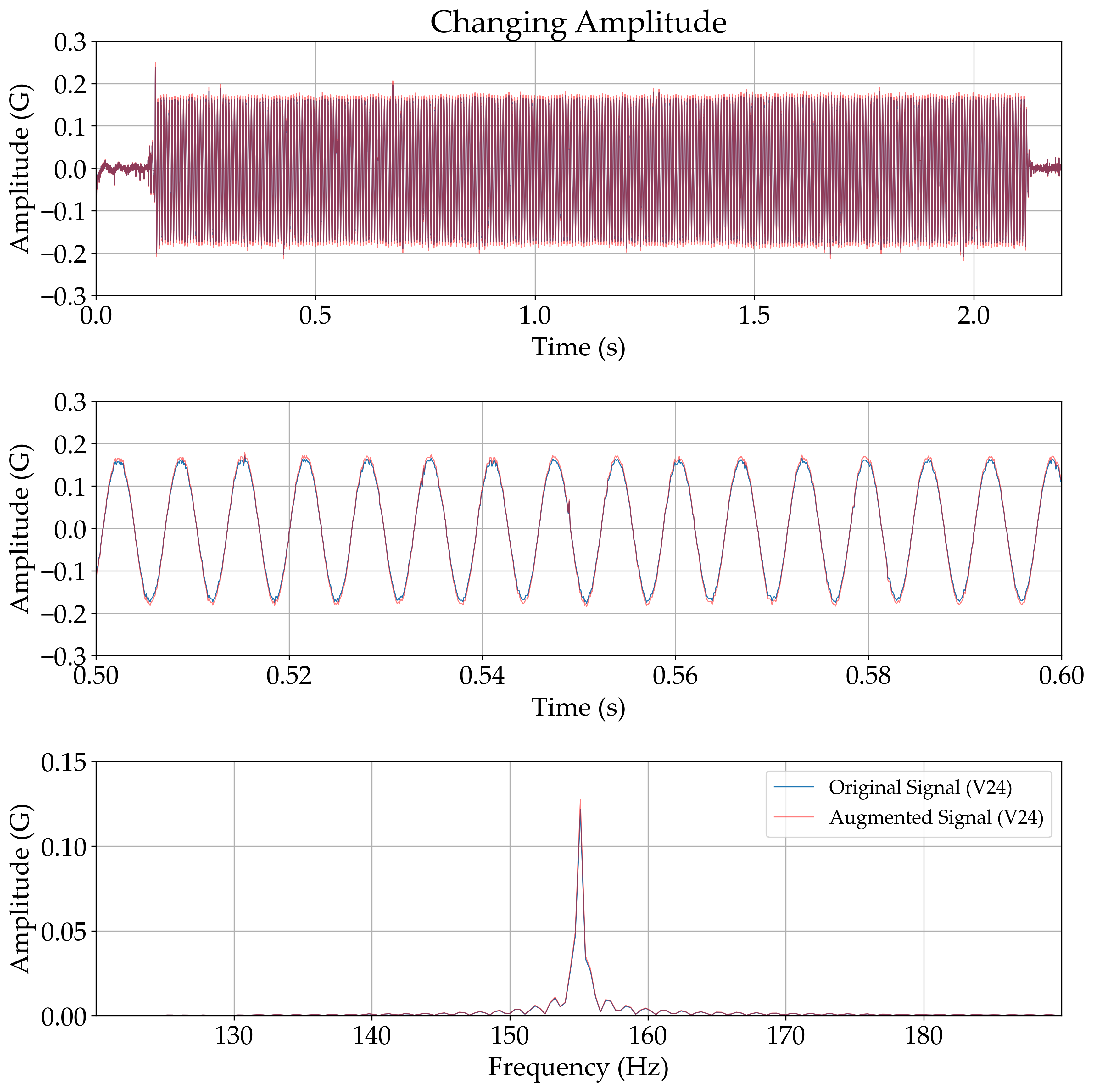}
    \label{fig:aug-amplitude}
    }

    \caption{
    Visualization of the three proposed augmentation techniques for mechanical vibrations, using V24 created with a carrier frequency of 155\,Hz.
    The rows, respectively, display the acceleration waveform over the entire duration, a zoomed-in view of the waveform between 0.5 and 0.6 seconds in the time domain, and the corresponding frequency domain plot using the Fast Fourier Transform (FFT).
    }
  \label{fig:augmentation}
\end{figure*}

In contrast to the extensive research and techniques available for data augmentation in images or audio, limited techniques exist for augmenting haptic data.
To address this, we propose three vibration augmentation techniques informed by human haptic perception and audio data augmentation methods, particularly for waveforms in the time domain, while ensuring that users cannot distinguish between the original and augmented vibrations (Figure~\ref{fig:augmentation}).
Five users confirmed that accelerations augmented by the following three techniques and their four combinations (=$\binom{3}{2}$+$\binom{3}{3}$) were indistinguishable through 3AFC (Three-Alternative Forced Choice) testing, where participants were presented with two identical stimuli (A, A; both being the original vibration) and one different stimulus (B; the augmented vibration) and asked to identify the different one.
For the seven augmentation cases, the average probability of participants correctly identifying the augmented vibration converged to around 0.33, suggesting the perceptual invariance of the augmented vibrations.

\subsubsection{Noise Injection}
We injected white noise into the original haptic data, ensuring that the noise intensities remained below the absolute detection thresholds or absolute limens (AL) for vibration frequencies (Figure~\ref{fig:aug-noise}).
Since humans are most sensitive to vibrations at around 200\,Hz~\cite{choi2012vibrotactile}, we injected uniform white noise with an amplitude $a$ within the range of [$-$0.0006\,G, 0.0006\,G].
The value 0.0006\,G represents the AL at 200\,Hz, and $a$ was randomly selected from this range.

\subsubsection{Changing Speed}
We increased or decreased the speed of the vibration, taking into account the discrimination thresholds or just noticeable differences (JND) for vibration frequency in haptics and the effects of duration on vibration perception, particularly temporal summation and amplitude JND (Figure~\ref{fig:aug-frequency}).
Since human frequency JND generally cluster between 15\% and 30\%~\cite{choi2012vibrotactile}, we set the change in the duration of the haptic data to be lower than 15\%.
Besides the frequency JND, since vibration durations longer than 10\,ms affect both the JNDs of vibration amplitude and the perceived intensity of vibration, we changed the speed of the haptic data at a rate $b$ within the range of $[-15\%, 15\%]$, where $b$ was randomly selected from this range, while ensuring that the absolute change in duration between the original and augmented haptic data was less than 10\,ms.

\subsubsection{Changing Amplitude}
We increased or decreased the amplitude of the vibration, considering the amplitude JND (Figure~\ref{fig:aug-amplitude}).
Since human amplitude JND mostly fall between 10\% and 30\%~\cite{choi2012vibrotactile}, we set the change in the amplitude of the haptic data to be lower than 10\%.
Therefore, we changed the amplitude of the haptic data at a rate $c$ within the range of $[-10\%, 10\%]$, where $c$ was randomly selected from this range.

\subsubsection{Implementation of Haptic Data Augmentation}
Based on the three augmentation techniques described above and their four combinations, we applied a total of seven augmentation methods to each of our 5,544 acceleration data, repeating the process twice.
In other words, we generated an additional 77,616 acceleration data entries ($= 5,544 \times 7 \times 2$) from the original 5,544 data, resulting in a total of 83,160 data entries.
We then fed them into the mechanoreceptive processing and the neural network.

\subsection{Mechanoreceptive Processing}
\label{sec:mechanoProcessing}
To better capture human vibrotactile processing, we supplemented the 1D waveform haptic data, which contains only amplitude values and lacks explicit frequency information, with additional data in a different form.
Previous research on developing neural networks for audio data introduced a method to convert the audio classification/prediction problem into an image classification/prediction problem by transforming waveform audio data into a spectrogram~\cite{shen2018natural, meng2019speech}.
This conversion can complement the waveform data by segmenting its duration into smaller time intervals and representing the frequency content of these segments in a single plot through the STFT.
In addition, since the converted spectrogram is an image, it is well-suited for input into CNN-based architectures designed for image processing.

When processing the conversion, we applied two bandpass filters based on the spectral sensitivity of mechanoreceptors involved in coding touch information for perception.
In vision research, neural networks are typically developed using three-channel (RGB) images~\cite{goodfellow2016deep, krizhevsky2012imagenet}, as the cone photoreceptors in the retina, which are responsible for visual processing, are classified into three types, each exhibiting different spectral sensitivities and therefore activated by photons of specific wavelengths~\cite{bowmaker1980visual}.
Similarly, each of the four types of mechanoreceptors involved in tactile processing have distinct spectral sensitivities and activation patterns.
However, we used two bandpass filters corresponding to the spectral sensitivities of Meissner Corpuscle (RA1) [3\,Hz, 100\,Hz] and Pacinian Corpuscle (RA2) [40\,Hz, 500\,Hz], as these two mechanoreceptors are most relevant to vibration perception~\cite{johnson2001roles}.
In addition, we also tested neural networks using four mechanoreceptive filters but did not find significant improvements in performance compared to those using two filters (see Section~\ref{sec:results} and Table~\ref{tab:comparison}).

We applied the mechanoreceptive processing technique to 83,160 augmented 1D waveforms, creating two-channel 2D spectrograms with raw floating-point numbers, where time is on the x-axis and frequency on the y-axis.
To ensure consistency in data sample length, we first added zero padding to the end of each acceleration data to make all lengths identical at 6,000 samples (i.e., 6-second durations), accounting for the maximum duration of the Tactons in our dataset, which varied in duration.
We then applied STFT to the two signals from the above mechanoreceptive filters, using a 0.5-second window and a 0.05-second hop size to achieve a spectral resolution of 2\,Hz in the spectrograms.
As a result, we derived two-channel spectrograms with dimensions of (251, 121), and then fed both the 83,160 1D acceleration data entries and their corresponding two-channel spectrograms into the neural network.

\section{VibNet: A Neural Network for Predicting Sensory and Emotional Ratings}
In this section, we propose a deep learning (DL) model, VibNet, to predict sensations and emotions conveyed through Tactons on a scale from 0 to 100.
VibNet consists of two parallel streams of neural networks inspired by human vibrotactile perception and cognitive mechanisms~\cite{kandel2000principles, lim2023can}, each utilizing different input types: 1D waveform data and two-channel 2D spectrograms, respectively.
By extracting both sequential and temporal-spectral features from these varied inputs, our proposed DL model aims to enhance the ability to predict the sensory and emotional impacts of haptic feedback.

\begin{figure*}[t]
  \centering
    \includegraphics[width=\linewidth]{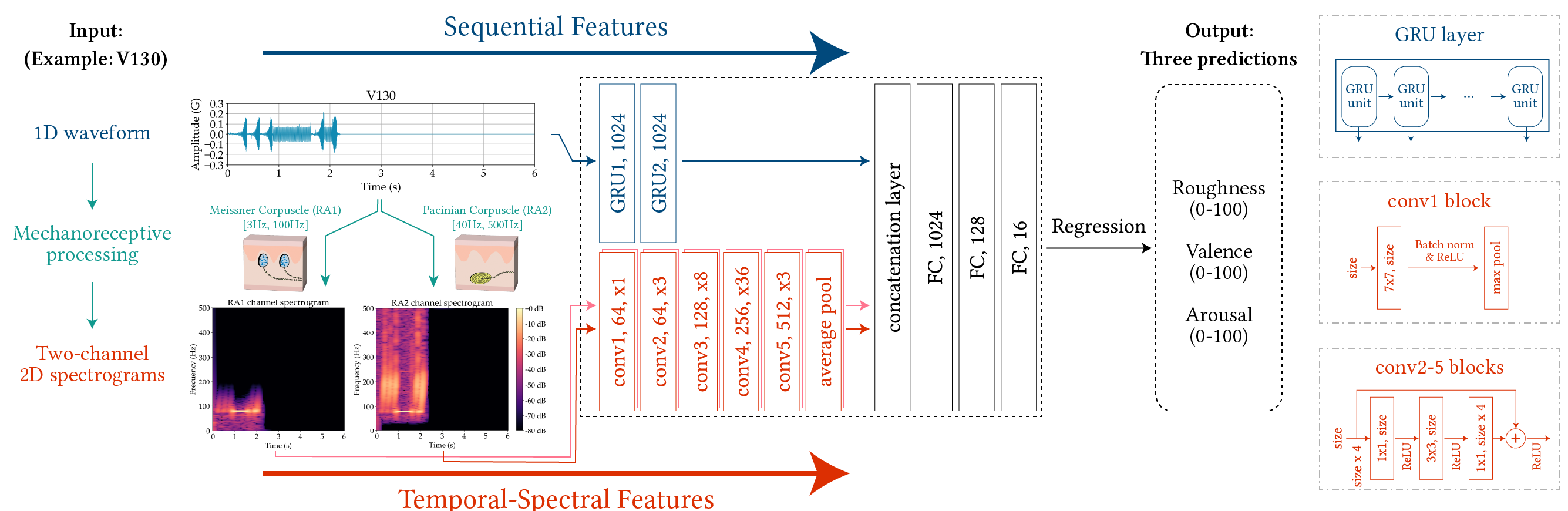}

    \caption{
      Overview of mechanoreceptive processing and VibNet's architecture, which consists of two parallel neural network streams.
    }

  \label{fig:architecture}
\end{figure*}

\subsection{Network Architecture}
VibNet consists of two parallel neural network streams, each designed to capture different aspects of the vibration data (Figure~\ref{fig:architecture}).
The first stream employs Gated Recurrent Units (GRU) layers~\cite{cho2014learning}.
GRU networks use a reset gate and an update gate in each unit to capture temporal dependencies and are known for their training efficiency compared to Long Short-Term Memory (LSTM) networks, as GRUs have fewer parameters.
Given the significant impact of temporal and rhythmic structures on human vibrotactile perception~\cite{lim2023can, park2011perceptual, ternes2008designing}, we selected GRU networks to effectively capture the sequential features of Tactons.
This stream consists of two layers, each with 1024 GRU units.
The first layer processes the 1D waveform data as input and passes it to the next layer.
The final GRU layer flattens the output before passing it to the concatenation layer.

The second stream employs a ResNet architecture~\cite{he2016deep}, a type of Convolutional Neural Network (CNN), to process the two spectrograms from each acceleration signal.
ResNet uses residual connections to allow the training of very deep networks by addressing the vanishing gradient problem, and it has demonstrated state-of-the-art performance with fewer parameters compared to traditional plain CNNs, such as VGGNet~\cite{simonyan2014very}.
We used a 154-layer ResNet with bottleneck blocks to capture the temporal-spectral features from the two-channel 2D spectrograms, effectively transferring the model's capabilities from image processing to haptic data processing.
As in the first stream, the final layer of this stream flattens the output and passes it to the concatenation layer.

The concatenation layer receives the outputs from the final layers of both streams.
This is followed by three fully connected layers with 1024, 128, and 16 neurons, respectively, applying the ReLU activation function and a dropout rate of 0.5 between layers.
The final layer, consisting of 16 neurons, regresses the values to provide three outputs corresponding to predictions for roughness, valence, and arousal ratings.

\subsection{Implementation}
We implemented VibNet using the PyTorch library, and trained it on the abovementioned haptic dataset using an NVIDIA GeForce RTX 3080 Ti GPU.
We employed the Adam optimizer with a batch size of 32 and used Mean Square Error (MSE) as the loss function.
The learning rate was set to 0.001, and the model was trained for 100 epochs, following commonly-used values in DL~\cite{goodfellow2016deep}.
To ensure robust performance, we applied 5-fold cross-validation during the training process, balancing training efficiency and model accuracy given the complexity of the dataset and architecture.

\section{Evaluation}
We evaluated the performance and generalizability of our framework using five baseline machine learning models on two new Tacton sets.
The baseline models were trained using the same haptic dataset as our model, with adjustments to meet the specific input requirements of each baseline.

\subsection{Method}
We verified VibNet against five baseline models, including conventional machine learning models and standard neural networks in the haptics field~\cite{awan2023predicting, gao2016deep}.
Additionally, we conducted an ablation study~\cite{sheikholeslami2019ablation} to assess the impact of the multi-channel spectrogram inputs and the corresponding CNN layers derived from the proposed mechanoreceptive processing (Section~\ref{sec:mechanoProcessing}). 
We compared the proposed framework — which utilizes two-channel spectrograms processed by two mechanoreceptive filters (RA1 and RA2) — with neural networks that applied (1) STFT without mechanoreceptive filters and (2) STFT with four mechanoreceptive filters (RA1, RA2, SA1, and SA2) in the second stream constructed by CNN layers.
We specifically tested the approach (2) because Merkel Disk (SA1) and Ruffini Ending (SA2) are known to contribute less to vibration perception compared to the Meissner Corpuscle (RA1) and Pacinian Corpuscle (RA2)~\cite{johnson2001roles, choi2012vibrotactile}.
To include these two additional channels, we applied a lowpass filter at 5\,Hz and a bandpass filter of [15\,Hz, 400\,Hz] considering the spectral sensitivities for Merkel Disk and Ruffini Ending.

We used two new Tacton sets from a previous study~\cite{lim4785071emotional}, which included sensory and emotional ratings for 24 Tactons in each set.
The first set of 24 Tactons was designed using sinusoidal parameters, while the second set consisted of 24 Tactons with complex waveforms.
None of the 48 Tactons in the two sets overlapped with the 154 Tactons used for training.
We collected acceleration data from 36 new participants for the 48 Tactons, following the same procedure described in Section~\ref{sec:setup}.
We compared the RMSE averaged across the 36 participants for roughness, valence, and arousal ratings between the predictions and the ground truths.
We also report the proportion of our predictions that fall within the standard deviation of user ratings across the two test sets.
We used individual acceleration data from each of the 36 participants, with each data entry consisting of 48 vibration measurements, to evaluate the model's performance for variations in user measurements.
However, we averaged the roughness, valence, and arousal ratings across participants and used these as the ground truth to ensure the results were generalizable and applicable to real-world haptic design, reflecting aggregate perceptions across diverse users.

\begin{table*}
\caption{
Comparison of RMSE between VibNet, the five baselines, and the two additional approaches of our framework.
The acronyms R, V, and A represent roughness, valence, and arousal, respectively.
}
\begin{tabular}{|c|cc|cccc|cccc|c|}
\hline
\multirow{2}{*}{Method}                                                     & \multicolumn{2}{c|}{\multirow{2}{*}{Network}}                                                                                                                                           & \multicolumn{4}{c|}{\begin{tabular}[c]{@{}c@{}}Test Set 1\\ (24 Tactons)\end{tabular}}                    & \multicolumn{4}{c|}{\begin{tabular}[c]{@{}c@{}}Test Set 2\\ (24 Tactons)\end{tabular}}                    & \multirow{2}{*}{Avg.} \\ \cline{4-11}
                                                                            & \multicolumn{2}{c|}{}                                                                                                                                                                   & \multicolumn{1}{c|}{R}         & \multicolumn{1}{c|}{V}         & \multicolumn{1}{c|}{A}         & Avg.      & \multicolumn{1}{c|}{R}         & \multicolumn{1}{c|}{V}         & \multicolumn{1}{c|}{A}         & Avg.      &                       \\ \hline \hline
\multirow{5}{*}{Baseline}                                                   & \multicolumn{2}{c|}{Linear Regression}                                                                                                                                                  & \multicolumn{1}{c|}{17.86}     & \multicolumn{1}{c|}{15.38}     & \multicolumn{1}{c|}{19.49}     & 17.58     & \multicolumn{1}{c|}{16.55}     & \multicolumn{1}{c|}{14.41}     & \multicolumn{1}{c|}{20.11}     & 17.02     & 17.30                 \\ \cline{2-12} 
                                                                            & \multicolumn{2}{c|}{1D-CNN~\cite{gao2016deep}}                                                                                                                                                      & \multicolumn{1}{c|}{17.94}          & \multicolumn{1}{c|}{11.14}          & \multicolumn{1}{c|}{17.37}          & 15.48          & \multicolumn{1}{c|}{23.00}          & \multicolumn{1}{c|}{11.54}          & \multicolumn{1}{c|}{22.60}          & 20.05          & 17.77                      \\ \cline{2-12} 
                                                                            & \multicolumn{2}{c|}{LSTM~\cite{gao2016deep}}                                                                                                                                              & \multicolumn{1}{c|}{18.19}          & \multicolumn{1}{c|}{9.44}          & \multicolumn{1}{c|}{18.69}          &          15.44 & \multicolumn{1}{c|}{17.13}          & \multicolumn{1}{c|}{9.01}          & \multicolumn{1}{c|}{18.71}          &        14.95
 & 15.20          \\ \cline{2-12} 
                                                                            & \multicolumn{2}{c|}{CNN-LSTM~\cite{awan2023predicting}}                                                                                                                                                      & \multicolumn{1}{c|}{17.71}     & \multicolumn{1}{c|}{8.81}      & \multicolumn{1}{c|}{17.90}     & 14.81     & \multicolumn{1}{c|}{16.61}     & \multicolumn{1}{c|}{8.16}      & \multicolumn{1}{c|}{16.94}     & 13.90     & 14.36  
        \\ \cline{2-12} 
                                                                            & \multicolumn{2}{c|}{Transformer~\cite{vaswani2017attention}}                                                                                                                                                      & \multicolumn{1}{c|}{21.28}     & \multicolumn{1}{c|}{12.90}      & \multicolumn{1}{c|}{20.13}     & 18.10     & \multicolumn{1}{c|}{18.47}     & \multicolumn{1}{c|}{9.39}      & \multicolumn{1}{c|}{19.08}     & 15.64     & 16.87  
                                                                            \\
                                                                            \hline \hline
\multirow{3}{*}{\begin{tabular}[c]{@{}c@{}}\\Proposed\\ Method\end{tabular}} & \multicolumn{1}{c|}{\multirow{3}{*}{\begin{tabular}[c]{@{}c@{}}\\Tested\\ Approach\end{tabular}}} & Single Spectrogram                                                                         & \multicolumn{1}{c|}{16.12}          & \multicolumn{1}{c|}{8.31}          & \multicolumn{1}{c|}{16.45}          & 13.63          & \multicolumn{1}{c|}{14.14}          & \multicolumn{1}{c|}{6.78}          & \multicolumn{1}{c|}{14.52}          & 11.81          & 12.72                      \\ \cline{3-12} 
                                                                            & \multicolumn{1}{c|}{}                                                                            & \begin{tabular}[c]{@{}c@{}}\textbf{Two-Channel}\\ \textbf{Spectrograms}\\ \textbf{(VibNet)}\end{tabular} & \multicolumn{1}{c|}{\textbf{15.12}} & \multicolumn{1}{c|}{\textbf{7.46}} & \multicolumn{1}{c|}{\textbf{14.86}} & \textbf{12.48} & \multicolumn{1}{c|}{\textcolor{blue}{\textbf{13.53}}} & \multicolumn{1}{c|}{\textcolor{blue}{\textbf{6.65}}} & \multicolumn{1}{c|}{\textcolor{blue}{\textbf{13.97}}} & \textcolor{blue}{\textbf{11.35}} & \textcolor{blue}{\textbf{11.91}}             \\ \cline{3-12} 
                                                                            & \multicolumn{1}{c|}{}                                                                            & \begin{tabular}[c]{@{}c@{}}Four-Channel\\ Spectrograms\end{tabular}                  & \multicolumn{1}{c|}{\textcolor{blue}{14.60}}          & \multicolumn{1}{c|}{\textcolor{blue}{7.21}}          & \multicolumn{1}{c|}{\textcolor{blue}{14.56}}          & \textcolor{blue}{12.12}          & \multicolumn{1}{c|}{13.60}          & \multicolumn{1}{c|}{7.61}          & \multicolumn{1}{c|}{14.22}          & 11.81          & 11.97                      \\ \hline
\end{tabular}
\label{tab:comparison}
\end{table*}

\subsection{Results}
\label{sec:results}
The RMSE values averaged across 36 users showed that all three of our proposed methods, regardless of whether mechanoreceptive filters were used in the second stream, outperformed the five baselines across both test sets (Table~\ref{tab:comparison}).
Among all tested models, linear regression yielded the highest RMSE for test set 1 (17.58).
The 1D-CNN model~\cite{gao2016deep} showed a reduction in RMSE for test set 1 but had the highest RMSE for test set 2 (20.05) among all tested models.
The LSTM model~\cite{gao2016deep} performed similarly to the 1D-CNN model for set 1 but showed improved performance for set 2, with a reduced RMSE of 14.95.
The CNN-LSTM network from~\cite{awan2023predicting}, which used parallel streams of CNN and LSTM, performed better than using either a single stream of CNN or LSTM~\cite{gao2016deep}.
However, the transformer regression model~\cite{vaswani2017attention} did not yield accurate predictions for the roughness and emotions of unseen Tactons, despite being well-trained.
Our proposed methods outperformed the CNN-LSTM network across all three tested approaches.
Within these approaches, using two mechanoreceptive filters led to a lower RMSE for test set 1 (RMSE = 12.48) compared to using a single spectrogram (i.e., without a mechanoreceptive filter, RMSE = 13.63).
The average RMSE for test set 1 was higher when using two mechanoreceptive filters compared to four mechanoreceptive filters (RMSE = 12.12).
For test set 2, the RMSE was lowest when using two mechanoreceptive filters (RMSE = 11.35) compared to using a single spectrogram or four-channel spectrograms (RMSE = 11.81).
Overall, VibNet with the two-channel mechanoreceptive filters performed the best over both datasets.

On average, 82\% of the predictions generated by VibNet fell within the standard deviation of the ground truth user ratings in both test sets across 36 participants.
Specifically, in set 1, 18.5, 23.6, and 16.9 out of 24 predictions for roughness, valence, and arousal, respectively, were within the standard deviation.
In set 2, 19.4, 23.6, and 16.2 out of 24 predictions for roughness, valence, and arousal, respectively, met this criterion.
Overall, across the two test sets, the highest proportion of predictions within the ground truth's standard deviation was for valence (98\% for both sets 1 and 2), followed by roughness (set 1: 77\%, set 2: 81\%) and arousal (set 1: 70\%, set 2: 67\%).




\section{Discussion}
In this paper, we designed 154 Tactons and ran a user study to construct a large-scale haptic dataset.
We augmented the acceleration data considering the human perception threshold and verified the perceptual invariance of the augmented 1D haptic signals.
The vibration signals were processed into two-channel spectrograms inspired by the spectral sensitivities of mechanoreceptors and fed into VibNet, which consisted of two streams of recurrent neural network layers and CNN layers.
The results demonstrated state-of-the-art performance on two unseen Tacton sets compared to the five baseline machine learning models.
Based on these findings, we discuss the efficacy of VibNet and outline implications for future work.

\subsection{Performance Comparison in Vibrotactile Sensation and Emotion Prediction}
We used dual streams in our neural network to capture sequential and temporal-spectral features, drawing inspiration from human vibrotactile perception and the cognitive mechanisms~\cite{kandel2000principles, lim2023can}.
Our evaluation demonstrated that this architectural choice enabled VibNet to achieve the lowest average RMSE across the test sets.
The 1D CNN model~\cite{gao2016deep} performed better than linear regression for test set 1, which consisted of Tactons designed with sinusoidal parameters that varied the spectral aspects of vibration signals.
However, this model showed the worst performance for test set 2, which consisted of Tactons with complex waveforms in the time domain.
In contrast, the CNN-LSTM model~\cite{awan2023predicting}, which combined 1D CNN and LSTM in dual streams to predict haptic attributes of real textures, showed lower RMSEs for the test sets compared to using a single stream alone.
Our findings further confirm that predicting vibration cognition requires capturing both sequential and temporal-spectral features, in line with the results presented in~\cite{lim2023can}.
Moreover, the results imply that using 2D spectrograms with 2D CNN layers instead of 1D waveforms with 1D CNN layers can enhance the network's ability to capture the spectral features of vibrations.

In the second stream designed to capture the temporal-spectral features of Tactons, the use of two-channel mechanoreceptive filters — based on the spectral sensitivities of the Meissner Corpuscle (RA1) and Pacinian Corpuscle (RA2) — led to a lower average RMSE of 11.91 across the two test sets compared to two alternative approaches: (1) using a single spectrogram (RMSE = 12.72) and (2) using four-channel mechanoreceptive filters (RMSE = 11.97). 
Specifically, while using more channels in the spectrogram improved performance for test set 1, the best performance for test set 2 was achieved with VibNet (RMSE = 11.35), followed by both the single spectrogram and the four-channel mechanoreceptive filters (RMSE = 11.81 for both).
These results underscore the efficacy of mechanoreceptive filters while also highlighting the need to balance the two streams within the network to maintain generalizability for predicting vibrotactile sensations and emotions. 
Overall, the predictions generated by VibNet achieved an average generalization accuracy of 82\% across Tacton sets and participants, as measured by the proportion of predictions falling within the standard deviations of the ground truths, irrespective of using any individual participant's acceleration haptic data.

\subsection{Limitations}
While VibNet demonstrated promising prediction performance and generalizability across different Tacton sets and users, our biomimetic framework has several limitations.
Although we trained VibNet on a wide range of Tactons designed using a variety of parameters, the model may be less accurate on Tactons outside the design space used for training. 
For instance, the training data included Tactons with durations ranging from 300 milliseconds to 6 seconds.
If a user inputs a Tacton lasting 20 seconds, VibNet may require adjustments to maintain accuracy.
Similarly, the frequency bandwidth of iPhones is in the range of 80\,Hz to 230\,Hz, with a maximum intensity around 0.3\,G.
Thus, making reliable predictions for Tactons outside these frequency and amplitude parameters may require additional data collection and further enhancements to the neural network.
Lastly, our focus was solely on haptic stimuli, with no auditory or visual stimuli provided.
Future research should aim to expand the dataset to cover a broader design space and incorporate multimodal stimuli to further enhance VibNet's utility.

\subsection{Implications for Future Work}
We outline how our framework can inform future research and influence haptic design practices.

\textbf{Designers can utilize VibNet to estimate the user's sensations and emotions when prototyping Tactons.}
Exploring the sensations and emotions elicited by Tactons is often a time-consuming and resource-intensive process, as it requires conducting multiple user studies while navigating an extensive Tacton design space.
However, VibNet requires only acceleration data produced by a haptic device, enabling designers to efficiently prototype and refine Tactons to achieve the desired sensations and emotions.
By reducing reliance on trial-and-error prototyping and extensive user testing, VibNet streamlines the development of haptic applications.

\textbf{Researchers can integrate our model into Tacton design tools.}
Designing effective Tactons for real-world applications is often a complex and challenging task. 
To address this, previous studies have proposed graphical user interfaces to assist designers in creating Tactons for diverse applications and scenarios~\cite{schneider2016studying, paneels2013tactiped}.
VibNet can further enhance these design tools by enabling designers to quickly evaluate the sensations and emotions elicited by the created Tactons, leading to a more efficient and streamlined design process.

\textbf{Our framework can inform the development of future machine learning models to predict various attributes conveyed by haptic stimuli.}
First, we provide a unique haptic dataset consisting of acceleration data collected from 36 participants, along with roughness, valence, and arousal ratings for 154 Tactons designed using a wide range of design parameters.
This dataset serves as a resource for training future prediction models in haptics domain.
Additionally, researchers can use this data to explore which components of neural network architectures contribute to capturing specific features of vibrations, providing deeper insights for Tacton design.
Second, our framework offers a method for augmenting haptic data offline while ensuring perceptual invariance of the signals. 
This method provides a foundation for developing deep learning or generative models that require large haptic datasets.
Researchers could also explore the effects of haptic data augmentation on model robustness, performance, and training efficiency, as has been done in vision and auditory research~\cite{simard2003best, ko2015audio}.
In addition, they could develop online augmentation techniques, such as using Generative Adversarial Networks (GANs)~\cite{goodfellow2020generative}, similar to those employed in vision and auditory research~\cite{ko2015audio, cubuk2019autoaugment}.
Such advancements could accelerate the development of computational models for haptics by enabling real-time data generation and refinement.
Finally, our demonstration of mechanoreceptive processing and dual-stream neural networks offers a foundation for future prediction models of Tactons. 
Future work can expand these models to predict specific metaphors conveyed through Tactons, such as heartbeat or tapping, with the collection of metaphor ratings~\cite{kwon2023can}.
Researchers could also extend VibNet to predict sensations and emotions elicited by vibrotactile grids~\cite{schneider2015tactile, israr2011tactile} or by multi-modal feedback systems, including thermal feedback and force feedback.
Moreover, future studies could expand the scope of prediction models beyond mechanical vibrations to encompass other stimuli in emerging haptic technologies, such as surface electrovibrations or mid-air ultrasound vibrations. 
Such models would not only improve in versatility and applicability but also enhance their relevance to a broader range of haptic interfaces and experiences.

\section{Conclusion}
The models for predicting vibrotactile sensations and emotions offer new opportunities for designers and researchers.
We proposed a framework grounded in the perception and cognition mechanisms of human haptic processing.
We hope our framework will help designers and researchers quickly prototype rich and diverse haptic feedback to convey target roughness and emotions, accelerating the integration of haptics into various user applications.

\bibliographystyle{IEEEtran}
\bibliography{Bibliography}

\begin{IEEEbiography}[{\includegraphics[width=1in,height=1.25in,clip,keepaspectratio]{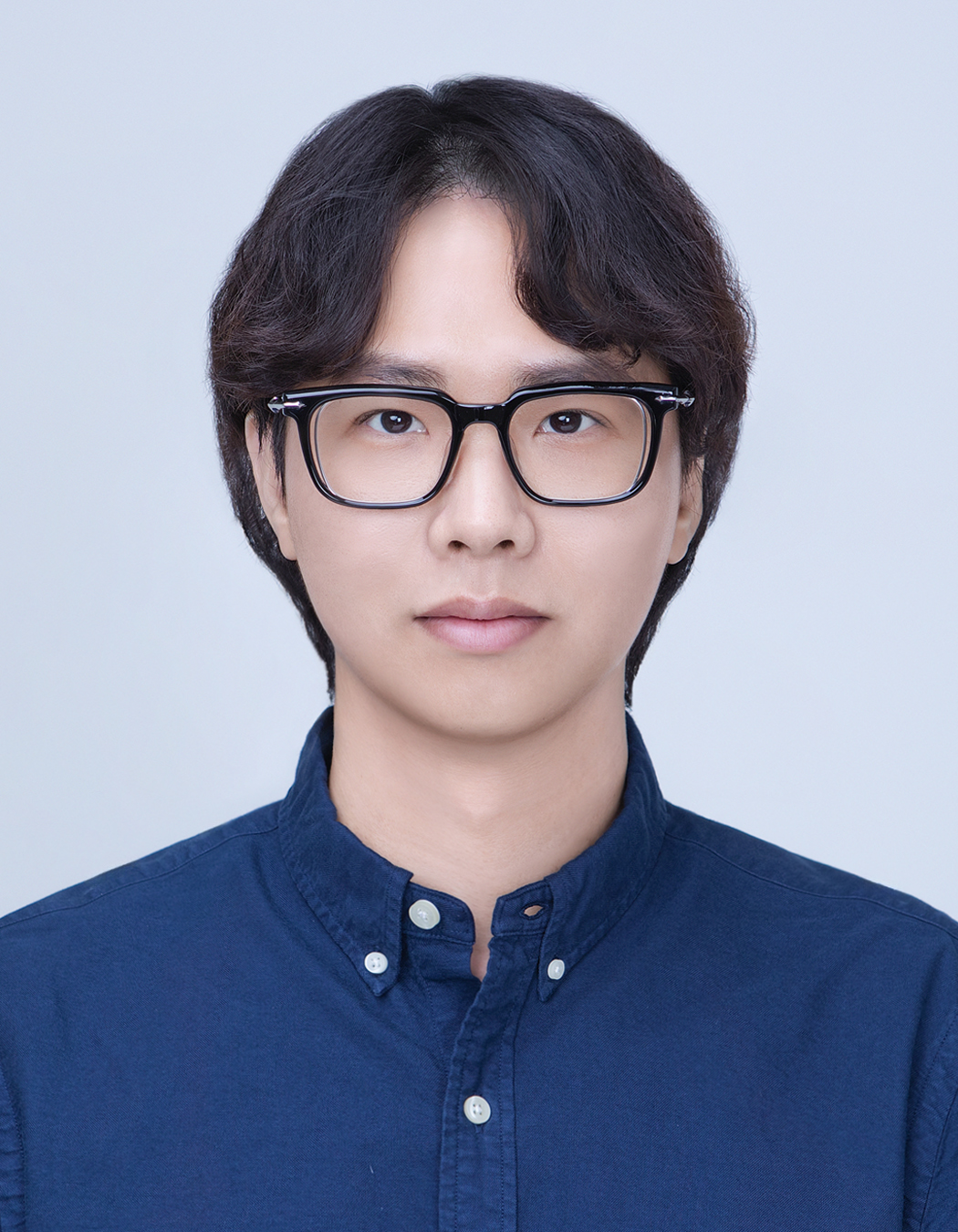}}]{Chungman Lim}
received the B.S. degree in statistics from the University of Seoul. He is currently a Ph.D. candidate in the Artificial Intelligence Graduate School at the Gwangju Institute of Science and Technology (GIST). His research interest is at the intersection of human-computer interaction, computational interaction, and haptics.
\end{IEEEbiography}

\begin{IEEEbiography}[{\includegraphics[width=1in,height=1.25in,clip,keepaspectratio]{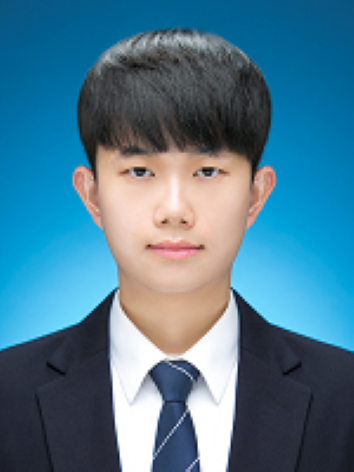}}]{Gyeongdeok Kim}
received the BS degrees both in mechanical engineering and computrer science engineering from Korea University of Technology and Education, Cheonan, South Korea, in 2020. He is a PhD student at Gwangju Institute of Science and Technology, Gwangju, South Korea.
His research interests include accessible system for the visually impaired, haptics, and human-computer interaction.
\end{IEEEbiography}

\begin{IEEEbiography}[{\includegraphics[width=1in,height=1.25in,clip,keepaspectratio]{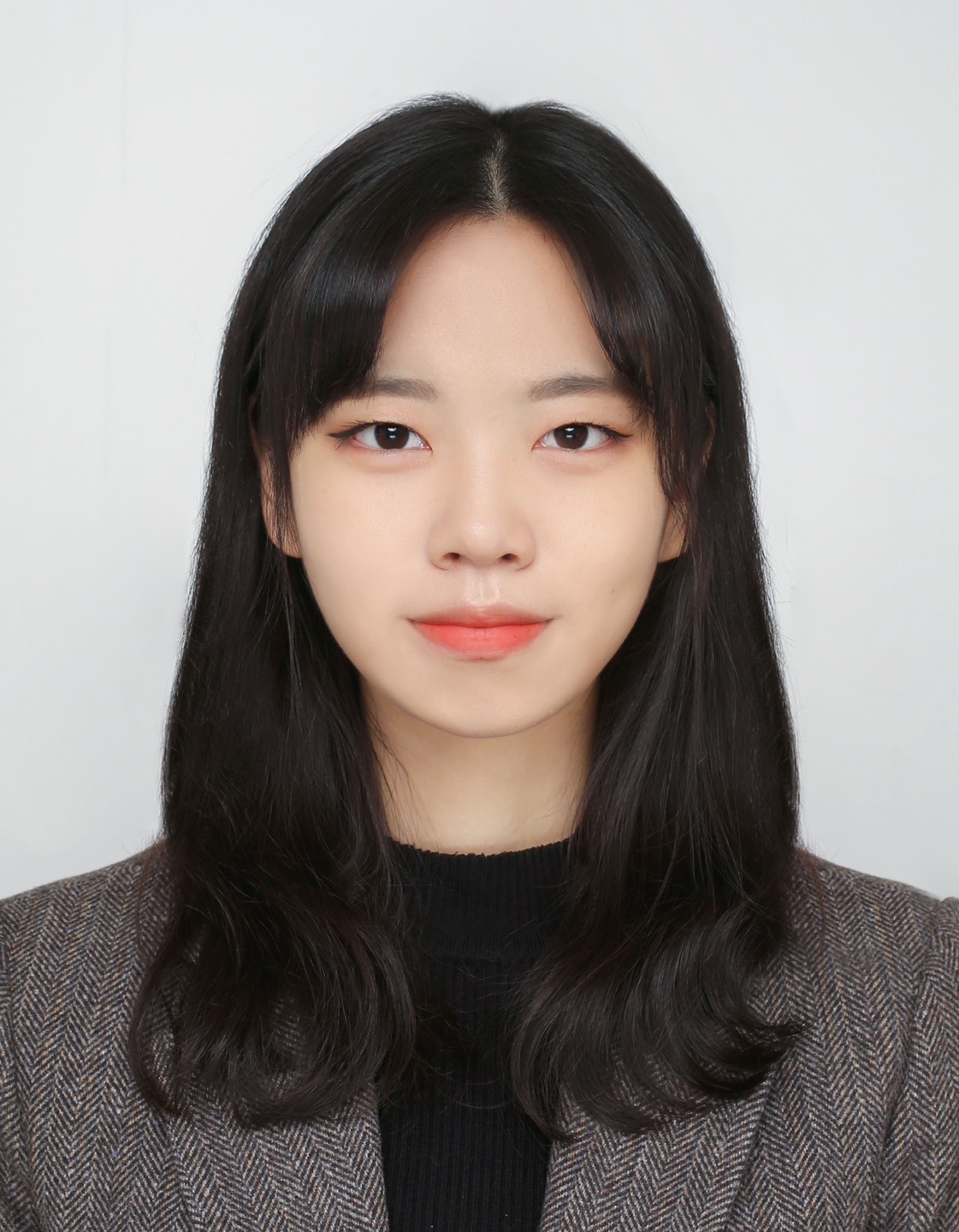}}]{Su-Yeon Kang}
received the BS degree in Computer Science and Engineering from Gyeongsang National University, Jinju, South Korea, in 2024. She is a master's student at the Gwangju Institute of Science and Technology, Gwangju, South Korea. Her research interests include human-computer interaction, haptics, and Artificial Intelligence.
\end{IEEEbiography}

\begin{IEEEbiography}[{\includegraphics[width=1in,height=1.25in,trim={3cm 5cm 4cm 1cm},clip,keepaspectratio]{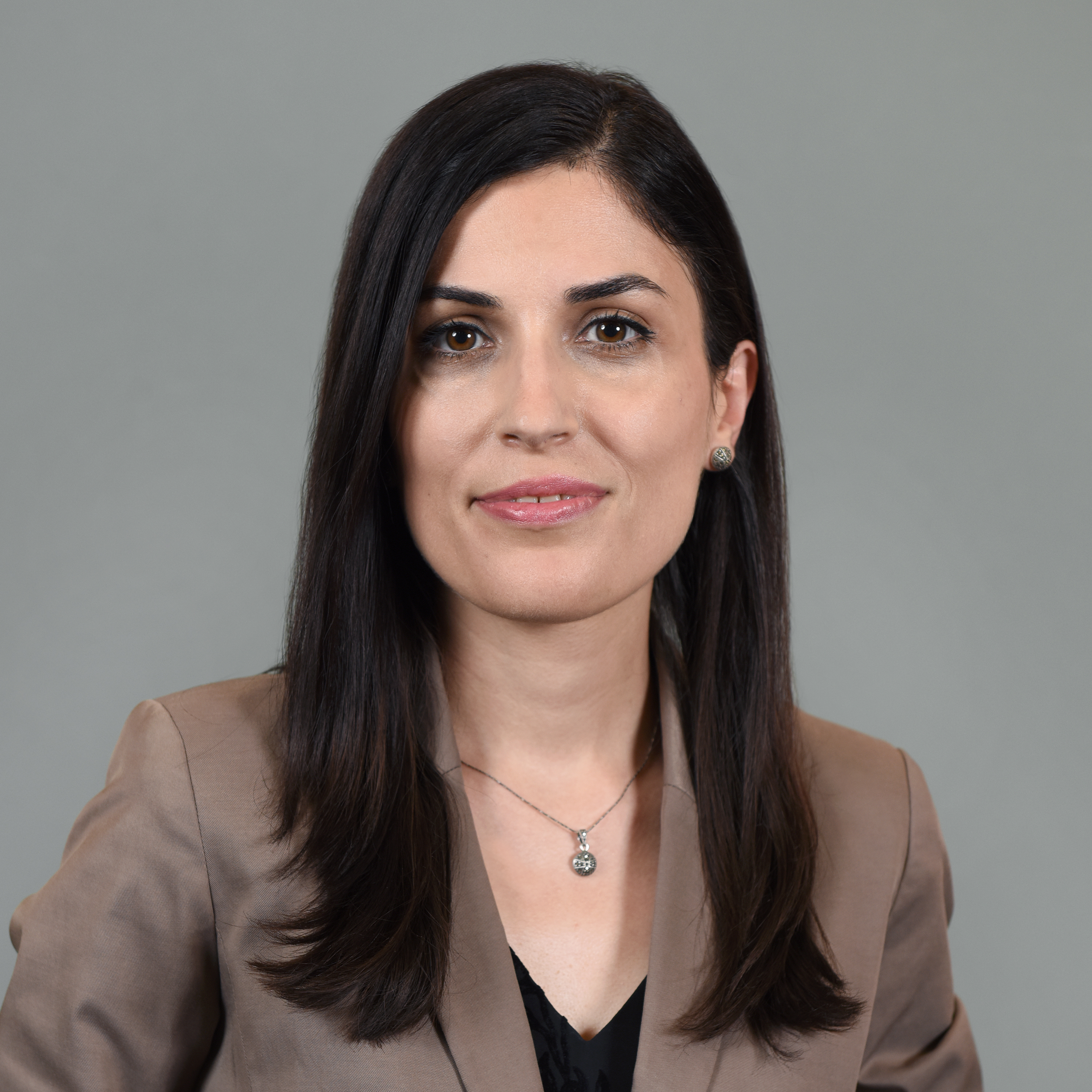}}]{Hasti Seifi} is an assistant professor in the School of Computing and Augmented Intelligence at Arizona State University. Previously, she was an assistant professor at the University of Copenhagen (2020-2022) and a postdoctoral research fellow at the Max Planck Institute for Intelligent Systems (2017-2020). She received her Ph.D. in computer science from the University of British Columbia in 2017, her M.Sc. from Simon Fraser University in 2011, and B.Sc. from the University of Tehran in 2008. Her research interests lie at the intersection of haptics, human-computer interaction, social robotics, and accessibility. Her work was recognized by an NSF CAREER award (2024), an NSERC postdoctoral fellowship (2018), and the EuroHaptics best Ph.D. thesis award (2017). 
\end{IEEEbiography}

\begin{IEEEbiography}[{\includegraphics[width=1in,height=1.25in,clip,keepaspectratio]{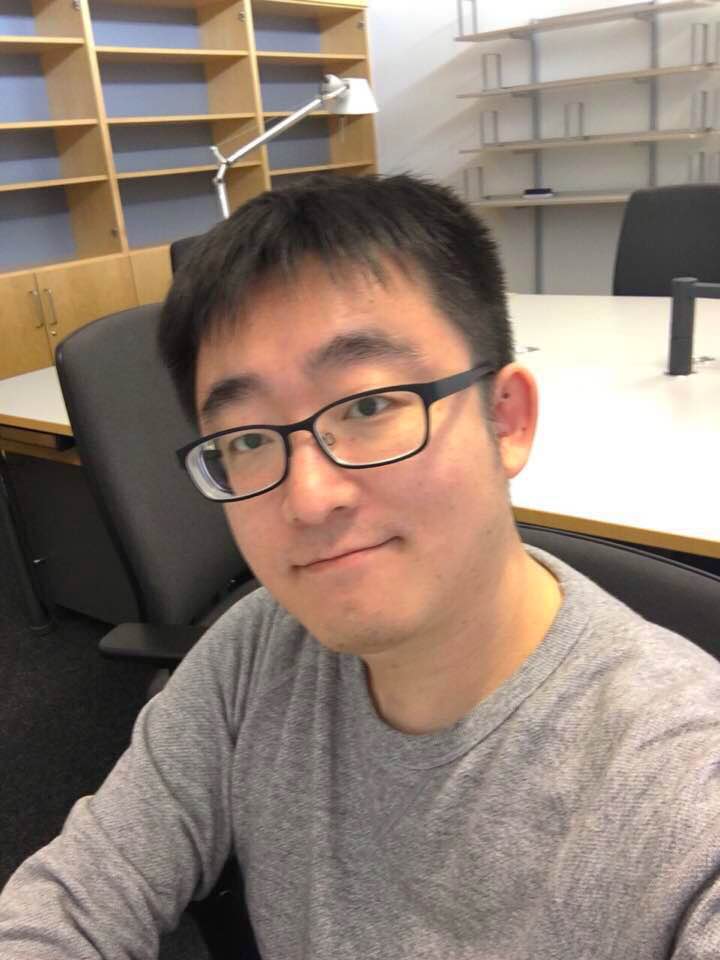}}]{Gunhyuk Park} received the B.S. degrees in computer science and engineering and in electronic and electrical engineering in 2007, and the Ph.D. degree in computer science and engineering in 2017 from the Pohang University of Science and Technology, Pohang, South Korea. He is currently an Assistant Professor at the AI Graduate School, Gwangju
Institute of Science and Technology, Gwangju, South Korea. He was a Postdoctoral Researcher with the Max Planck Institute for Intelligent Systems from 2017 to 2019. His research interests include haptic rendering and perception, mainly on tactile systems. He has worked on haptics for vibrotactile feedback perception, illusory vibrotactile feedback, automobiles, and mobile phones.
\end{IEEEbiography}


\vfill

\end{document}